\documentclass[floats,floatfix,showpacs,amssymb,prd,twocolumn,superscriptaddress,nofootinbib,longbibliography,reprint]{revtex4-2}
\usepackage{amssymb,amsmath,verbatim,mathtools,needspace,enumitem,etoolbox,graphicx,physics,microtype,afterpage,xspace,tabularx,lmodern,multirow}
\usepackage{gensymb}
\usepackage{placeins}
\usepackage[dvipsnames, usenames]{xcolor}
\definecolor{linkcolor}{rgb}{0.0,0.3,0.5}
\usepackage[unicode, colorlinks=true, linkcolor=linkcolor, citecolor=linkcolor, filecolor=linkcolor, urlcolor=linkcolor, linktocpage, breaklinks]{hyperref}
\usepackage[all]{hypcap}
\usepackage[T1]{fontenc}
\usepackage[utf8]{inputenc}
\usepackage{mathrsfs}
\usepackage[usenames,dvipsnames]{xcolor}

\usepackage[normalem]{ulem}
\hypersetup{colorlinks=true,citecolor=romared,linkcolor=romared,urlcolor=romared}
\definecolor{romared}{RGB}{142,0,28}
\usepackage{aas_macros}
\usepackage{makecell}
\usepackage{soul}
\usepackage[nolist,nohyperlinks]{acronym}
\usepackage{orcidlink}
\usepackage{academicons}
\usepackage{lipsum}

\definecolor{orcidlogocol}{HTML}{A6CE39}
\usepackage{mathtools}

\newcommand{\orcid}[1]{\href{https://orcid.org/#1}{\includegraphics[width=10pt]{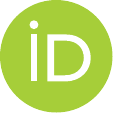}}}
\newcommand{\lm}{\ell m}
\newcommand{\Cross}{\mathbin{\tikz [x=1.4ex,y=1.4ex,line width=.2ex] \draw (0,0) -- (1,1) (0,1) -- (1,0);}}%
\newcommand{\Energy}{\mathcal{E}}
\newcommand{\AngMom}{\mathcal{L}}
\newcommand{\Carter}{\mathcal{Q}}
\newcommand{\costh}{\mathsf{z}}
\newcommand{\ddx}[2]{\frac{\mathrm{d} #1}{\mathrm{d} #2}}

\newcommand{\radialPot}{\mathcal{R}}
\newcommand{\polarPot}{\mathcal{Z}}
\newcommand{\rroot}{\mathsf{r}}

\newcommand{\be}{\begin{equation}}
\newcommand{\ee}{\end{equation}}
\newcommand{\beq}{\begin{eqnarray}}
\newcommand{\eeq}{\end{eqnarray}}
\newcommand{\Quantity}[3]{\ensuremath{\Bigl(\frac{#1}{#2 \, #3}\Bigr)}}

\graphicspath{{./Figures/}}

\allowdisplaybreaks

\begin{document}

\pagenumbering{arabic}

\title{Relativistic aerodynamics of spinning black holes}
\author{Conor Dyson \orcid{0000-0002-9742-9422}}
\email[]{conor.dyson@nbi.ku.dk}
\affiliation{Niels Bohr International Academy, Niels Bohr Institute, Blegdamsvej 17, 2100 Copenhagen, Denmark}
\author{Jaime Redondo-Yuste \orcid{0000--0003--3697--0319}}
\email[]{jaime.redondo.yuste@nbi.ku.dk}
\affiliation{Niels Bohr International Academy, Niels Bohr Institute, Blegdamsvej 17, 2100 Copenhagen, Denmark}
\author{Maarten van de Meent \orcid{0000-0002-0242-2464}}
\affiliation{Niels Bohr International Academy, Niels Bohr Institute, Blegdamsvej 17, 2100 Copenhagen, Denmark}
\affiliation{Max Plank Institute for Gravitational Physics, Am Mühlenberg 1, 14476 Potsdam, Germany}
\author{Vitor Cardoso \orcid{0000--0003--0553--0433}}
\affiliation{Niels Bohr International Academy, Niels Bohr Institute, Blegdamsvej 17, 2100 Copenhagen, Denmark}
\affiliation{CENTRA, Departamento de F\'{\i}sica, Instituto Superior T\'ecnico -- IST, Universidade de Lisboa -- UL,
Avenida Rovisco Pais 1, 1049--001 Lisboa, Portugal}
\affiliation{Yukawa Institute for Theoretical Physics, Kyoto University, Kyoto
606-8502, Japan}
\date{\today}

\begin{abstract}
Astrophysical black holes do not exist in vacuum, and their motion is affected by the galactic environment. As a black hole moves it attracts stars and matter, creating a wake that, in turn, exerts an effective friction slowing down the black hole. This force is known as dynamical friction, and has significant consequences, ranging from the formation of supermassive black hole binaries to modifications in the phase of binary mergers. In this work we explore the motion of {\it spinning} black holes on a medium. We find that the classical ``drag'' along the velocity direction is modified and two novel forces appear: a rotational force, which in the context of fluid dynamics is dubbed the Magnus force, and a lift, orthogonal to the direction of motion. 
We develop a first--principles fully--relativistic treatment of these spin-induced aerodynamic forces in two types of environment: i) collisionless corpuscular matter and ii) a light scalar field, exploring the differences between both cases. In both cases we find that the total rotational force acts precisely in the opposite direction as compared to the classical set--up of a spinning ball moving through a fluid.
Finally, we comment on the consequences of these new effects for astrophysics and gravitational wave observations.  
\end{abstract}

\maketitle

\section{Introduction}\label{sec:Introduction}
When a small body passes by a more massive object it can experience an acceleration, known as the slingshot effect, a purely gravitational phenomena. This net transfer of momentum must then decelerate the massive companion. If this process happens repeatedly, for example, when a massive star moves through the galactic medium, it produces significant friction on the motion of the object. This effect was dubbed \emph{dynamical friction}, since the net result is a force parallel and in the opposite direction of the velocity of the object~\cite{chandrasekhar1943dynamical, Chandrasekhar:1943ys}. An alternative viewpoint is that in the frame of the massive object, it tends to pull the smaller objects towards itself. However since this object is moving, the smaller objects will cluster at some distance behind the actual position of the object, forming a \emph{wake}. The gravitational pull that this wake exerts on the object is directly related to the dynamical friction force~\cite{danby1957statistical, kalnajs1971polarization, mulder1983dynamical}.

The impact of dynamical friction on galactic dynamics is ubiquitous~\cite{Milosavljevic:2001vi,Milosavljevic:2002bn, DelPopolo:2003ap, Boylan-Kolchin:2007bvo}. It causes proto--planets growing in disks to slow down, and migrate towards the center of solar systems. A similar effect happens in stellar clusters, forcing the more massive stars towards the center. Black holes (BHs) are also subject to this effect: in particular dynamical friction could be the key to bringing together supermassive BH binaries after galactic mergers, and could be a mechanism used by unequal-mass binaries to ``swim'' across a galaxy or accretion disk~\cite{Begelman:1980vb, Milosavljevic:2002bn, Shannon:2015ect, Cardoso:2021wlq, Cardoso:2020nst}.

The occurrence of dynamical friction does not rely on the specific properties of the medium. It exists whether the medium is composed of collisionless particles (such as a star moving through galactic gas, planets and smaller stars), a homogeneous dark matter halo made of axions~\cite{Ostriker:1998fa, Weinberg:1978abc, Abbott:1982af, Preskill:1982cy, Dine:1982ah, Hui:2016ltb,Vicente:2022ivh}, or, generally, ultralight bosonic particles~\cite{Arvanitaki:2009fg, Brito:2015oca, Irastorza:2018dyq}. Thus, dynamical friction may provide a measurable imprint of dark matter, by affecting the emission of gravitational waves in compact binary coalescences~\cite{Macedo:2013qea,Annulli:2020lyc,Coogan:2021uqv,Cole:2022yzw,Santoro:2023lek,Tomaselli:2023ysb}. Moreover, understanding dynamical friction in wave--like mediums is of relevance to model the dynamics of BHs in accretion disks (of another, potentially supermassive BH)~\cite{Duffell:2019uuk,Derdzinski:2018qzv,ONeill:2024tnl}, or in dense clusters. For these reasons, characterising dynamical friction at all possible wavelengths, from the small wavelength limit (point--particle case), to the regime where the wavelength is comparable, or even larger than the BH itself (ultralight dark matter) is crucial to model a variety of astrophysical scenarios~\cite{Barausse:2014tra,Kocsis:2011dr}.

In recent years, there have been several attempts at providing a first principle calculation of dynamical friction in the fully relativistic, wave--like regime~\cite{Hui:2016ltb, Annulli:2020lyc, Vicente:2022ivh}, as well as efforts from the numerical relativity (NR) point of view~\cite{Traykova:2021dua, Aurrekoetxea:2023fhl}.  In this endeavor, good agreement has been obtained between both approaches~\cite{Vicente:2022ivh,Traykova:2023qyv} in those regimes where they are expected to be comparable. 

An important difference between BHs and other massive objects such as stars or planets is that astrophysical BHs have relativistic spins which leads to an axially rather than spherically symmetric gravitational potential. Indeed, astrophysical BHs that grow through accretion might not only be spinning, but potentially doing so very rapidly~\cite{Thorne:1974ve}. It stands to reason that we should study carefully how the spin of a BH affects dynamical friction. By breaking the cylindrical symmetry of the problem, it is natural to expect novel effects that appear when the spin of the BH is not aligned with its direction of motion. In this work we will explore this in detail.

We will show that due to the breaking of cylindrical symmetry, the momentum transferred to the BH will no longer be necessarily parallel to the BHs velocity vector. This spin--dependent force can be decomposed into an orthogonal triad of forces, which we will refer to collectively as the aerodynamic forces. The dominant contribution is the well--studied drag force, whose direction is parallel to the BH velocity. This is always the larger effect since objects that are far away from the BH do not ``feel'' its rotation, but are still sensitive to its Newtonian gravitational pull. Second to this is a force that is reminiscent of the Magnus effect in fluid dynamics~\cite{2012PrAeS..55...17S}: this force acts in a direction which is orthogonal to the plane spanned by the BHs velocity and its direction of rotation
\footnote{Previous works have shown that given the total force orthogonal to the spin and velocity vectors a division can be made into individual terms, one of which taking the explicit form $J\Cross S$, which they argue is the appropriate quantity to refer to as the Magnus force~\cite{Costa:2018gva}. In addition, they also note that when this term is separated out it carries the same characteristic sign as the classical Magnus force. This is a subtle and interesting point of which readers should be aware. In this work, however, we simply define the Magnus to be the total force orthogonal to the spin and velocity vectors, aligned with the direction of the classical Magnus, disconnecting it somewhat from the classical picture.\label{footnote_costa}}.
Finally, the BH also suffers a lift (or a downforce) in a direction orthogonal to the direction of motion, but that lies in the plane spanned by the BH spin and velocity. Previous works \cite{Okawa:2014sxa, Frolov:2024xyo} have analysed the motion of Kerr BH through massless scalar and electromagnetic fields and slow--motion, weak--gravity descriptions were also worked out~\cite{Cashen:2016neh, Costa:2018gva}. In this work, we present an extensive, self-consistent, and fully relativistic description of the forces that arise in this setup. Moreover, we compute and compare our results between two different regimes: one where the medium is composed of 
collisionless massive particles, and one where the medium is provided by an ultralight scalar field, focusing on the case where the wavelength of the field is comparable to the size of the BH.

The structure of the paper is as follows:  Sections~\ref{sec:particle_medium} and \ref{sec:wave_medium} present the theory necessary for calculating the aerodynamic forces in particle and wave-like mediums respectively. In section~\ref{sec:comparison} we compare our methods for calculating aerodynamic forces to previous results.
We discuss our results in section~\ref{sec:results}, including the dependence of the aerodynamic forces on the spin of the BH, its velocity, and the incidence angle. Moreover we also provide polynomial fits that accurately describe our results in the low spin, low velocity regime. We comment on the consequences of the spin dependence of the aerodynamic forces in several astrophysically relevant scenarios in section~\ref{sec:discussion}. Finally we conclude and summarize our results in section~\ref{sec:conclusions}.
In the following, we will make use of the mostly plus signature for the metric, $\eta_{ab} = (-,+,+,+)$, lower case Latin indices representing abstract spacetime indices. A round bracket between a pair of indices denotes symmetrization of said pair. Unless otherwise specified we use geometric units where $G=\hbar=c=1$. 

\section{Particle--like environments}\label{sec:particle_medium}

We begin by studying the motion of a spinning BH in a medium of collision-less particles, with uniform density $\rho$, composed of particles with a mass scale $m_p$ distributed on a cloud with radius $b_{\rm max}$~\footnote{It is well known that the drag force is (logarithmically) divergent when the cloud size is taken to infinity, $b_{\rm max} \to \infty$~\cite{Chandrasekhar:1943ys, Vicente:2022ivh}, so imposing such a cut--off is necessary from the computational point of view.}. In the BH rest frame this problem is equivalent to scattering geodesics on that same BH background. We first obtain closed-form expressions for the transfer of momentum due to individual particles in geodesic motion. We numerically integrate the total contribution to the transfer of momentum due to each individual geodesic to compute the total force that the BH suffers. For this reason, although our approach is not fully analytical, it is sensible to all the strong--field effects. Further analytical developments can be carried within a post--Minkowskian approach. Recently, expansions of the scattering and inclination angles were obtained up to order $\mathscr{O}(G_N^3a^2)$, providing a potentially simpler integration scheme~\cite{Gonzo:2023goe}. However, obtaining fully analytic formulas for the drag, Magnus and lift forces in a slow--spinning post--Minkowskian approach is out of the scope of this work.

\subsection{Geodesics in the Kerr metric}
The exterior of a rotating BH is described by the Kerr metric, which only depends on the BH mass $M$ and its spin $J = a/M$. In Boyer--Lindquist coordinates the line element is given by 
\begin{equation}\label{eq:Kerr_Metric}
     \begin{aligned}
        ds^2 =& - \frac{\Delta}{\Sigma}\Bigl(dt - a\sin^2\theta d\phi\Bigr)^2 + \frac{\Sigma}{\Delta}dr^2 \\
        &+ \Sigma d\theta^2 + \frac{\sin^2\theta}{\Sigma}\Bigl(adt - (r^2+a^2)d\phi\Bigr)^2 \, ,
    \end{aligned}
\end{equation}
where 
\begin{equation}
    \Sigma = r^2 + a^2\cos^2\theta  \,  , \quad \Delta = r^2-2Mr+a^2    \,  .
\end{equation}
The geodesic equation for a massive particle, when parametrized by the Mino time $\lambda$ (satisfying $d\lambda= \Sigma d\tau$, where $\tau$ is the proper time)~\cite{Mino:2003yg}, are given by 
\begin{equation}\label{eq:Geodesic_Equations}
    \begin{aligned}
        \Bigl(\ddx{r}{\lambda}\Bigr)^2 =& \radialPot(r) \,  ,   \\
        \Bigl(\ddx{z}{\lambda}\Bigr)^2 =& \polarPot(z) \,     ,   \\
        \ddx{t}{\lambda} =& \frac{r^2+a^2}{\Delta}\Bigl(\Energy (r^2 + a^2) - a \AngMom\Bigr) - a^2 \Energy (1-z^2) + a \AngMom \, , \\
        \ddx{\phi}{\lambda} =& \frac{a}{\Delta}\Bigl(\Energy (r^2 + a^2) - a \AngMom  \Bigr) + \frac{\AngMom}{1-z^2} - a\Energy \, ,
    \end{aligned}
\end{equation}
with $z = \cos\theta$ and $\Energy, \AngMom, \Carter$ are the energy, the angular momentum, and the Carter constants of motion (see~\cite{Carter:1968rr, Fujita:2009bp} and references therein for the construction of the geodesic equations and their solutions in the bound and plunging scenarios). The exact form of the radial and polar (or angular) potentials $\radialPot, \polarPot$ are not relevant for this work, the only important feature being that $\radialPot$ is a polynomial of degree $4$, and $\polarPot\equiv \polarPot(z^2)$ is a polynomial of degree $2$ in $z^2$.
As a consequence, the radial potential has generically $4$ different roots, which we label $\rroot_i$ with $i = 1,\dots,4$, ordered in such a way that $1/\rroot_i \leq 1/\rroot_{i+1}$. Note that $r_1$ is negative for scattering orbits. Similarly we label the polar potential roots by $\costh_{1,2}$. Generic bound geodesics have two turning points $\rroot_{1(2)}$, which together with the polar turning point $\costh_1$ uniquely define the trajectory~\cite{Fujita:2009bp}. For instance, they directly define the eccentricy $e$, semi-latus rectum $p$, and maximum inclination $x_{\rm inc}$ via 
\begin{align}\label{eq:Eccentricity_Semilatus_From_Constants}
        p =& \frac{2\rroot_1\rroot_2}{\rroot_1+\rroot_2} \, , \\
        e =& \frac{\rroot_1-\rroot_2}{\rroot_1+\rroot_2} \, ,\\
        x_{\rm inc}^2 &= 1-z_1^2 \, . \label{eq:Inclination_From_Constants}
\end{align}
We are interested both in scattering geodesics~\footnote{These can be obtained via analytical continuation of the analytic solutions for bound geodesics obtained in~\cite{vandeMeent:2019cam} as explained in~\cite{vandeMeent:b2b}.} and plunging geodesics (which contribute through accretion onto the BH)~\cite{Dyson:2023fws}. The separatrix between both cases occurs when the second radial turning point becomes complex, $\rroot_2 = \rroot_3$. 
Equations~\eqref{eq:Eccentricity_Semilatus_From_Constants}--\eqref{eq:Inclination_From_Constants} allow us to describe equivalently the geodesic either knowing $\{p,e,x^2_{\rm inc}\}$, or the orbital constants $\{\Energy,\AngMom, \Carter\}$ (the radial and polar roots can be obtained analytically in terms of these, and vice--versa). 

In addition to the orbital constants, we need four initial phases $(q_t^0,q_r^0,q_z^0,q_\phi^0)$ to fully specify a geodesic~\cite{vandeMeent:2019cam}. Of these, we can always arrange $q_t^0$ and $q_\phi^0$ to be zero, by applying a time translation and rotation to the Kerr background. We can also choose to always start our Mino time parameter at the pericenter of the trajectory, giving $q_r^0=0$. For generic bound geodesics, we could have used ergodicity of the orbit to also set $q_z^0=0$ without lose of generality. This, however, is not possible for scattering geodesics, since they complete in a finite amount of Mino time~\cite{vandeMeent:b2b}. Different values of $q_z^0$ lead to physically different scattering orbits, and we will need to specify its value.

In order to compute the momentum flux imparted on the BH by an incoming particle, we make use of the scattering angle and the difference between the final and initial inclinations, which are~\cite{vandeMeent:b2b,Gonzo:2023goe}
\begin{equation}\label{eq:Inclination_Angles}
    \begin{aligned}
        \cos \theta_{\rm in/out} =& \mathsf{z}_1 \mathbf{sn}\Bigl[\mathsf{K}(k_z) \frac{2}{\pi}\Bigl(q_{z}^0 \mp \frac{\Upsilon_z}{\Upsilon_r} q_{r}^S\Bigr)\Big\vert k_z\Bigr] \, , \\
        k_z =& a^2 \Bigl(1-\Energy^2\Bigr) \frac{\mathsf{z}_1^2}{\mathsf{z}_2^2} \, , \\
        q_r^S =& \frac{\pi}{\mathsf{K}(k_r)}\mathsf{F}\Bigl[
        \arcsin\sqrt{\frac{\rroot_3-\rroot_1}{\rroot_2-\rroot_1}} \Big\vert k_r\Bigr] \, , \\
        k_r =& \frac{(\rroot_1-\rroot_2)(\rroot_3-\rroot_4)}{(\rroot_1-\rroot_3)(\rroot_2-\rroot_4)} \, .
    \end{aligned}
\end{equation}
Here, $\Upsilon_{z(r)}$ are polar (radial) frequencies given by 
\begin{equation}
    \begin{aligned}
        \Upsilon_r =& \frac{\pi}{2\mathsf{K}(k_r)} \sqrt{(1-\Energy^2)(\rroot_1-\rroot_3)(\rroot_2-\rroot_4)} \, , \\
        \Upsilon_z =& \frac{\pi \mathsf{z}_2}{2\mathsf{K}(k_r)} \, ,
    \end{aligned}
\end{equation}
$\mathsf{K}$ is the complete elliptic integral of the first kind, $\mathsf{sn}$ the Jacobi elliptic sine function, and $q_z^0$ is the initial polar phase. Similarly, the scattering angle can be obtained as 
\begin{equation}\label{eq:Scattering_Angle}
    \chi = 2q_r^S \frac{\Upsilon_z}{\Upsilon_r} - \phi_r(q_r^S) + \phi_z\Bigl(q_z^0 + q_r^S \frac{\Upsilon_z}{\Upsilon_r}\Bigr) - \phi_z(q_z^0 - q_r^S \frac{\Upsilon_z}{\Upsilon_r}\Bigr) \, ,
\end{equation}
where the phases $\phi_{r(z)}$ are defined in Eqs.(27)--(35) of~\cite{vandeMeent:2019cam}, and for simplicity we set the initial radial phase to zero. We compute these geodesic quantities using the Black Hole
Perturbation Toolkit \cite{BHPToolkit}.

\subsection{Forces from scattering geodesics}
%
\begin{figure}
\includegraphics[width = \columnwidth]{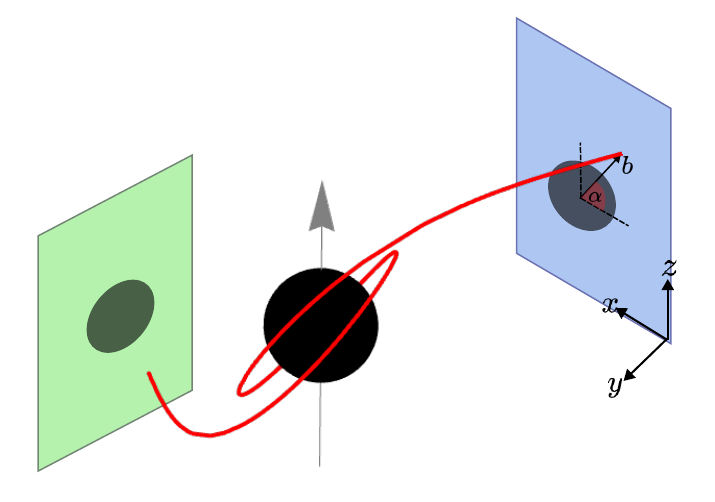}
\caption{Diagram representing the motion of a geodesic (red) and the coordinate system used to describe it, with $z$ pointing in the direction of the BH spin, and $y$ such that the particle is in-coming along the $(y,z)$ plane. The particle is identified by its impact parameter $b$ and impact angle $\alpha$, as shown in the blue impact plane. We will extract the momentum flux at the out--going plane, represented in green. }
\label{fig:Geodesics_Diagram}
\end{figure}
We will now compute the force that is imparted to the BH due to the absorption of particles of the medium (plunging geodesics) and conservation of linear momentum, associated to the scattering of geodesics. We consider a homogeneous medium composed of point--particles of density $\rho$ such that each individual mass $m_p \ll M$. The BH is moving with some velocity $v$ in a direction which forms an angle $\beta$ with its spin. For simplicity, we assume that the BH spin is oriented along the $z$ axis, see Fig.~\ref{fig:Geodesics_Diagram} for a illustration of our setup. The only free parameters in this case are the velocity $v$ of the BH and the angle $\beta$.  
In the BH frame, the motion of each particle can be mapped to either a scattering or a plunging geodesic. Each particle can be identified by the angle $\alpha$ that it forms with the plane defined by the BH spin, and a radial vector $b\in[0, b_{\rm max}]$ which can be understood as an impact parameter. This is illustrated in Fig.~\ref{fig:Geodesics_Diagram}. These variables provide a useful parametrization, from which we can define the particle's asymptotic four--velocity and impact vector as 
\begin{equation}\label{eq:Particles_Velocity_Impact}
    \begin{aligned}
        u^\mu =& \gamma \Bigl[1, 0, v \sin\beta, -v\cos\beta\Bigr] \, , \\
        b^\mu =& b \Bigl[0, \sin\alpha, -\cos\beta\cos\alpha, \sin\alpha, \sin\beta\cos\alpha\Bigr] \, .
    \end{aligned}
\end{equation}
Above, $\gamma = (1-v^2)^{-1/2}$ is the Lorentz factor. From these quantities, we can deduce the particle's constants of motion:
\begin{equation}
    \begin{aligned}
        \Energy =& \gamma \, , \\
        \AngMom =& -\gamma b v \sin\alpha \sin\beta \, , \\
        \Carter =& \frac{\gamma v^2}{4}\Bigl[2a^2-3b^2 + (2a^2-b^2)\cos(2\beta) \\
        &\qquad -2b^2\cos(2\alpha)\sin^2(\beta)\Bigr] \, .
    \end{aligned}
\end{equation}
As discussed previously, these three quantities define radial and polar roots of the particle's trajectory, which can be obtained in closed form (see e.g.~\cite{vandeMeent:b2b}). Once we know the roots, Eqs.~\eqref{eq:Inclination_Angles}--\eqref{eq:Scattering_Angle} determine the initial and final inclination angles, as well as the scattering angle $\chi$. The expression for the initial inclination angle is used to fix the initial phase $q_z^0$ such that $\theta_{\rm in} = \beta$.

Plunging geodesics correspond to particles that get accreted by the BH, and therefore, the total momentum transfer due to plunging geodesics, $\delta p^\mu_P$, by conservation of momentum, equals the initial momentum $m_p u^\mu_P$ of the plunging particles:
\begin{equation}
\frac{\delta p^\mu_{\rm P}}{m_p} = u^\mu_{\rm P} \, .
\end{equation}
This implies that we need to distinguish between scattering and plunging geodesics. As discussed above, the separatrix is found when the second and third radial roots coincide $\rroot_2=\rroot_3$. We solve numerically for this condition, building the separatrix $b_{\rm crit}(\alpha)$ in the impact plane for each configuration. Then, geodesics parametrized by some $\{b,\alpha\}$ such that $b \leq b_{\rm crit}(\alpha)$ plunge, whereas in the opposite case, they correspond to scattering orbits. 

\begin{figure}[t!]
    \includegraphics[width = \columnwidth]{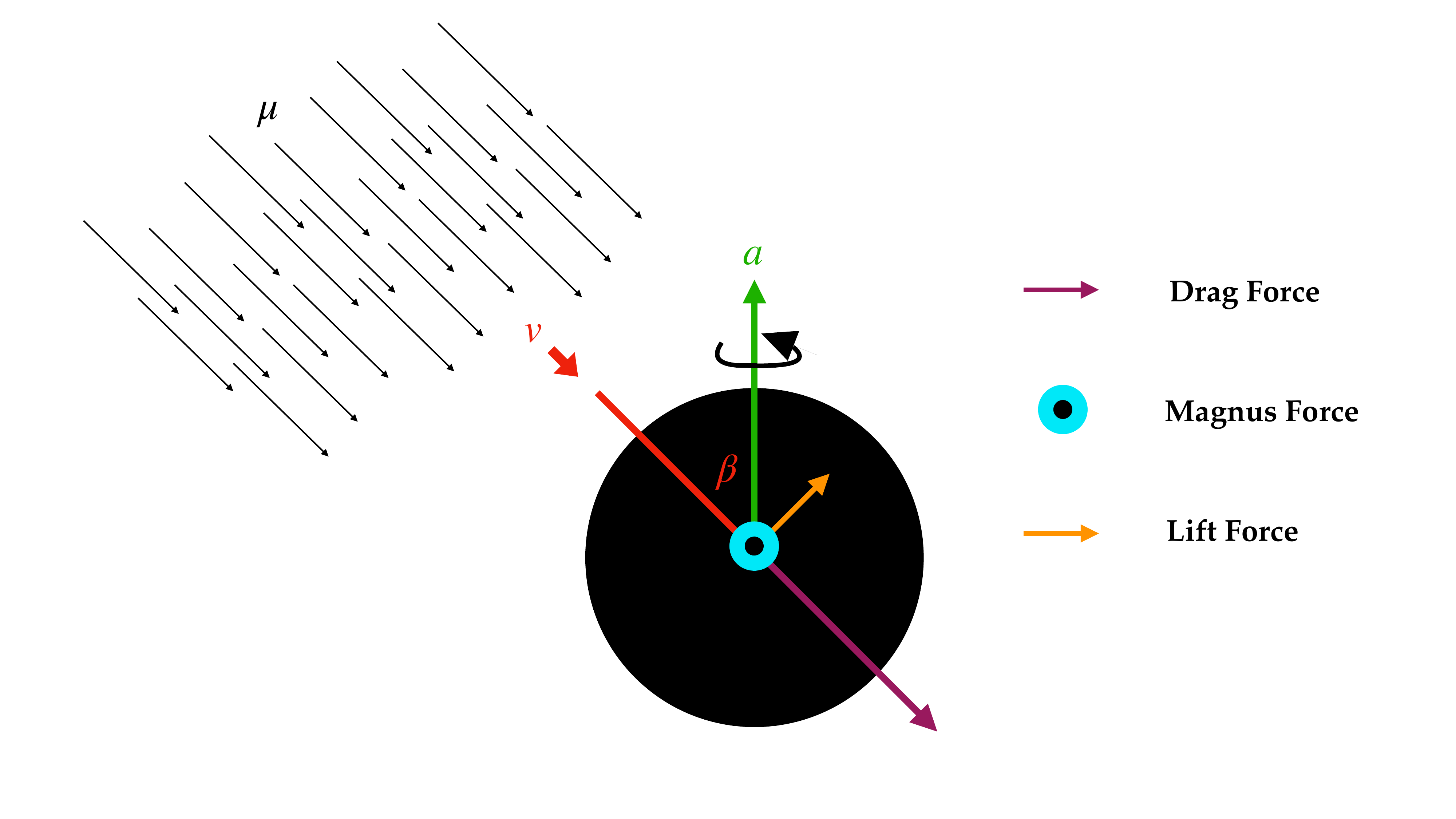}
    \caption{Schematic representation of the vector decomposition of the drag, Magnus, and lift forces relative to the velocity and spin directions. BH is viewed from side-on where in this frame the drag force is aligned with the velocity vector of the scalar field. The Magnus force in this representation is out of the plane and is given by the vector perpendicular to the plane spanned by the spin and velocity vectors (if the spin and velocity vectors are aligned then the Magnus force vanishes due to the symmetries of the system). The lift vector is in the direction perpendicular to the Magnus and drag vectors. Throughout this work, a positive sign for drag, Magnus and lift forces means they are aligned with the corresponding vectors in the figure.}
    \label{fig:forces}
\end{figure}

In the scattering regime, the momentum transfer can be directly computed once we know the final inclination $\theta_{\rm out}$ and the scattering angle $\chi$. Indeed, conservation of momentum yields
\begin{equation}
    \frac{\delta p^\mu_{\rm S}}{m_p} = \gamma v 
    \begin{pmatrix}
        0 \\
        \sin\theta_{\rm out}\sin\chi \\
        \sin\beta - \sin\theta_{\rm out}\cos\chi\\
        -\cos\beta - \cos\theta_{\rm out}
    \end{pmatrix} \, ,
\end{equation}
where we remind the reader that we have chosen the initial phase such that $\theta_{\rm in} = \beta$. We are not interested in the momentum transfer of each individual particle, but rather on the total momentum transfer due to the BH moving along a medium. We denote the total momentum transfer per unit mass by $\Delta \Tilde{p}^\mu$. This is given by 
\begin{equation}
    \begin{aligned}
        m_p \Delta \Tilde{p}^\mu =& \int_0^{2\pi}d\alpha \int_0^{b_{\rm crit}(\alpha)} b db \delta p^\mu_{\rm P}(b,\alpha) \\
        &+ \int_0^{2\pi}d\alpha \int_{b_{\rm crit}(\alpha)}^{b_{\rm max}} b db \delta p^\mu_{\rm S}(b,\alpha) \, .
    \end{aligned}
\end{equation}
In order to resolve accurately the second integral, the behavior of geodesics that end up scattering away, but whirl around the BH one or several times is crucial.  In light of this we found it convenient to change the coordinates to 
\begin{equation}
    b \mapsto b_{\rm crit(\alpha)} + \Bigl[b_{\rm max} - b_{\rm crit}(\alpha)\Bigr]e^{\Tilde{b}} \, ,
\end{equation}
and then integrate numerically between $\Tilde{b}\in[\Tilde{b}_{\rm min}, 0]$. We want to be able to resolve geodesics to sub--percent distance relative to the separatrix. Thus, we fix $\Tilde{b}_{\rm min} = -\log(10^{-\gamma}b_{\rm max})$, where $\gamma$ is the desired accuracy. For practical purposes we find that $\gamma= -3$ is enough to extract confidently the forces and achieve convergence. For convenience, we also change coordinates to resolve the plunge integral, defining $b = \Tilde{b}b_{\rm crit}(\alpha)$, so that the new variable is $\Tilde{b}\in[0,1]$.
From the momentum flux, the forces, as defined in Fig.~\ref{fig:forces} can be computed directly as 
\beq
F_D^{(P)} &=& \gamma v \rho M^2\Bigl( \Delta \Tilde{p}^z \cos\beta  - \Delta \Tilde{p}^y \sin\beta \Bigr) \, ,\label{F_dragp} \\
F_M^{(P)} &=& \gamma v \rho M^2\Delta \Tilde{p}^x \, ,\label{F_magnusp} \\
F_L^{(P)} &=& \gamma v\rho M^2 \Bigl(\Delta \Tilde{p}^y \cos\beta + \Delta \Tilde{p}^z \sin\beta \Bigr) \, , \label{F_liftp}
\eeq
where the $\gamma v M^{-1}$ factor is the dilation factor of the co--moving volume that the BH spans in a unit time. 
Recovering the mass units from $\Delta \Tilde{p}^\mu$ involves multiplying everything by the mass scale of the individual environment particles $m_p = \rho M^{3}$, which combined with the above yields the overall dimensionless $\rho M^{2}$ scaling factor. Our numerical results for forces will always be given in units of $\rho M^2$.
Finally, $F_{D,M,L}$ denote the drag, Magnus and lift forces, respectively.

The drag force is the usual dynamical friction force, which is antiparallel to the motion of the BH. The novel forces that appear when considering spinning BHs moving at an angle with respect to its spin axis are the Magnus and the lift forces. The Magnus force is normal to the plane defined by the BH spin and velocity directions. The lift force, on the other hand, is orthogonal to both the BH velocity vector and the Magnus force. This is represented schematically in Fig.~\ref{fig:forces}.  

Already at this level we can observe that plunging geodesics, i.e., those associated to particles that fall into the BH, do \emph{not} contribute to the Magnus or lift forces, resulting only in a net contribution to the drag due to accretion. Indeed, the force due to accretion is parallel to the BH velocity, therefore, its only contribution is in the usual dynamical friction~\cite{Cashen:2016neh}.

\section{Wave-like environments}\label{sec:wave_medium}

We now wish to model the aerodynamic forces in a wave-like medium. We begin by considering a minimally coupled massive scalar field~\cite{Klein:1926tv,Gordon:1926emj,Witek:2012tr}
\begin{equation}
    S = \int d^4 x \sqrt{-g} \left(\frac{R}{8 \pi} - \frac{1}{2}\nabla_a \Phi \nabla^a \Phi^* - \frac{1}{2}\mu^2 |\Phi|^2\right) \, , 
\end{equation}
where $\mu$ is the mass of the scalar field $\Phi$, and $\nabla$ denotes the covariant derivative. 
The equations of motion following from the action are
\begin{equation}\label{eq:EquationsOfMotion}
    \begin{aligned}
        \Box_g \Phi =& \mu^2 \Phi \, , \\
    R_{ab} - \frac{1}{2}R g_{ab} =& 8 \pi T_{ab} \,  ,
    \end{aligned}
\end{equation}
where $\Box_g$ is the d'Alembert operator associated to the metric $g$, $R_{ab}$ is the Ricci tensor, and the stress--energy tensor of the scalar field is given by
\begin{equation}\label{eq:stress}
T^{ab} = \nabla^{(a} \Phi \nabla^{b)} \Phi^* - \frac{g^{ab}}{2}  ( \nabla_{c}\Phi\nabla^{c}\Phi^* + \mu^2 |\Phi|^2 )\, .
\end{equation}
In the regime where back reaction of the scalar field on the metric is negligible, we can take $g_{ab}$ to be the Kerr metric. Then, substituting the solutions of $\Phi$ into Eq.~\eqref{eq:stress} we can use the stress--energy tensor to obtain asymptotic quantities relating to forces.

We will approach this problem in the frequency domain, and we expand the scalar field in spheroidal harmonics as
\begin{equation}
	\Phi = \sum_{\lm}  e^{- i \omega t} \phi_{\lm}(r)S_{\lm}(\theta, \varphi,\xi ) \, ,
\end{equation}
where  $S_{\lm}(\theta, \varphi, \gamma )$ are spheroidal harmonics with angular numbers $\lm$ and spheroidicty $\xi =  i k_{\infty}a$, with $J=aM$ being the angular momentum of the BH. Here we have defined the asymptotic wavenumber of the scalar field to be
\be
k_{\infty}=\sqrt{\omega^2 - \mu^2}\,.
\ee
The radial equation governing $\phi_{\lm}$ is~\cite{Starobinsky:1973aij}
\begin{align}\label{eq:radialeq}
\begin{aligned}
	&\Delta \frac{d}{dr}\left[\Delta\frac{d \phi_{\lm}}{dr}\right] + \bigg[ \omega^2(r^2+a^2)^2-4 a M m \omega r \\
		&\;\;\;\;\;\;+ (m a)^2-(\lambda_{\lm}+\mu^2(r^2+a^2))\Delta)\bigg]\phi_{\lm}=0.
\end{aligned}
\end{align}
Solutions of the previous equation that are regular asymptotically far away from the BH take the form
\begin{equation}\label{eq:asymptoticform}
	\phi_{\lm} \sim\frac{I_{\lm} e^{-i k_{\infty} r_*}+R_{\lm} e^{i k_{\infty} r_*  }}{\sqrt{r^2+a^2}} \, ,
\end{equation}
with $r_* = r + \eta \log{2 k_{\infty}r }$, and  $\eta \equiv M \left( \frac{\omega^2+k_{\infty}^2}{k_{\infty}^2} \right)$.
The asymptotic form of the field defines two independent solutions, $I_{\ell m}$ corresponding to in--going plane waves scattering towards the BH from past null infinity, $\mathcal{I}^-$, and $R_{\ell m}$ corresponding to out-going plane waves scattering off of the Kerr potential towards future null infinity, $\mathcal{I}^+$.

\subsection{Boundary conditions and field solution}

In order to solve Eq.~\eqref{eq:radialeq} and calculate asymptotic forces on the Kerr BH, we first consider some fixed background scalar field given by 
\begin{equation}\label{eq:backgroundscalar}
    \Phi = e^{-i \mu t}\, .
\end{equation}
This is a constant energy density solution to the Klein--Gordon equation in flat space, whose oscillation frequency is fixed by the mass of the field. It is also a valid solution to~\eqref{eq:EquationsOfMotion} at large radii, which is the physically relevant case for the cloud sizes we wish to consider here. By performing a Lorentz boost into the BH frame, the background field then takes the form
\begin{equation}
     \Phi_{BC} = e^{-i \mu \gamma t}e^{-i  \gamma\mu v r^{\rm out}_*(\cos\beta\cos\theta + \sin{\beta}\sin{\theta}\sin{\varphi})} \, ,
     \label{eq:phi_BC_arb}
\end{equation}
where $r_*^{\rm out} = r_*(r^{\rm out}_{\rm BC})$, and $r^{\rm out}_{\rm BC}$ is fixed as the radius at which we evaluate the plane wave boundary condition. Here we have also defined $v$ as the velocity of the BH, $\gamma$ as the Lorentz factor and $\beta$ as the inclination angle between the velocity vector and the spin--vector (which we fix to be the $z$-axis). The boost also redshifts the frequency of the field, fixing $\omega=\gamma\mu$, and $k_{\infty} = \gamma \mu v$.

We wish to set Eq.~\eqref{eq:phi_BC_arb} as an outer boundary condition to the piece of the field incoming from $\mathcal{I}^-$ in Eq.~\eqref{eq:radialeq}. In doing so we must first decompose Eq.~\eqref{eq:phi_BC_arb} in spheroidal harmonics. This is achieved through the following steps: we first rotate our coordinate system to a frame where $\beta=0$.
Next, we decompose the simplified form of Eq.~\eqref{eq:phi_BC_arb} with $\beta=0$ in spherical harmonics. After doing so we make use of the Wigner D-matrices to rotate back to the original frame with $\beta\neq 0$.
Finally, we obtain the spheroidal modes by summing the spherical decomposition from Eq.~\eqref{eq:phi_BC_arb} over spherical-spheroidal mixing coefficients.
We will begin with the spatial boundary condition in the rotated frame
\begin{equation}
	\widetilde{\Phi}_{BC} = e^{-i k_{\infty} r^{\rm out}_* \cos\theta}\label{eq:phi_BC}\, ,
\end{equation}
where the tilde denotes quantities defined in the rotated coordinate system. In this rotated system there is no $\varphi$ dependence, meaning we will only obtain non-zero axi--symmetric modes, i.e., modes with $m=0$.  

Next, we project this expression onto spherical harmonics,
\begin{equation}
    \begin{aligned}
        \tilde{\mathrm{W}}_{\lm} =& \int Y_{\lm} (\theta,\varphi) \tilde{\Phi}_{BC}  d \Omega \\
        =& 2 \pi \sqrt {\frac {2 \ell + 1}{4 \pi } }\delta_{0 m} \int_{-1}^{1}  \tilde{\Phi}_{BC}  P_{\ell}(x) dx \, ,
    \end{aligned}
\end{equation}
with $x=\cos\theta$. Using Rodrigues' formula we obtain an expression that can be integrated by parts $\ell$--times. All corresponding boundary terms vanish, leaving us with
\beq
	\tilde{\mathrm{W}}_{\ell \, m}(r) &=& (i k_{\infty}  r^{\rm out}_* )^\ell  \frac{ \sqrt{ (2 \ell + 1)\pi}}{2^\ell \Gamma(\ell+1)}\delta_{m 0}  \nonumber\\
 &\times&\int_{-1}^{1} e^{-i k_{\infty} r^{\rm out}_*  x} [(x^2-1)^\ell] dx \, ,
\eeq
which can be conveniently written in terms of Bessel functions as
\begin{equation}
\begin{aligned}
    \tilde{\mathrm{W}}_{\lm'} = & (-i)^{\ell} \pi\text{sgn}(k_{\infty}  r^{\rm out}_* ) \\ 
    &\times\sqrt{\frac{2(2\ell+1)}{k_{\infty}  r^{\rm out}_* }} J_{\ell+\frac{1}{2}}(k_{\infty}  r^{\rm out}_* )\delta_{m'0} \, .
 \end{aligned}
\end{equation}
Having obtained the decomposition in a simplified rotated frame, we can then appeal to the Wigner-D matrices to transform the spherical harmonics between different angular frames. Such that 
given $\tilde{\mathrm{W}}_{\ell m'}$ in a certain frame, then an equivalent $\mathrm{W}_{\lm}$ in a new rotated frame is given by
\begin{equation}
	\mathrm{W}_{\lm}  = \sum_{m'=-\ell}^{\ell} D^{\ell}_{mm'}(\alpha,\beta,\gamma)\tilde{\mathrm{W}}_{\ell m'},
\end{equation}
where $\alpha$, $\beta$ and $\gamma$ are the three Euler angles relating the two frames. In the context of our problem, we only require the $m'=0$ matrix 
\begin{equation}
 D^\ell_{m0} = \sqrt{\frac{(\ell-m)!}{(\ell+m)!} } P_{\lm}(\cos\beta) e^{-im\alpha}.
\end{equation}
The Euler angle $\alpha$ corresponds to rotations along the direction of the spin of the BH, which is an axis of symmetry in our problem, thus we set $\alpha = 0$  without loss of generality, yielding
\begin{equation}\label{eq:WLM}
\begin{aligned}
	\mathrm{W}_{\lm} =& (-i)^{\ell+m}  \pi \text{Sign}( k_{\infty} r^{\rm out}_* )  P_{\lm}(\cos\beta)\\
		&\times\sqrt{\frac{2(2\ell+1) (\ell-m)!}{k_{\infty} r^{\rm out}_* (\ell+m)!}} 
			J_{\ell+\frac{1}{2}}( k_{\infty} r^{\rm out}_* ).
\end{aligned}
\end{equation}
Here the $m$-mode dependence has now been reintroduced to our problem through the Wigner matrices and the full decomposition of the plane wave Eq.~\eqref{eq:phi_BC_arb} is now given by
\begin{equation}\label{eq:WLM}
\begin{aligned}
	\Phi_{BC} = \sum_{\ell, m} W_{\lm} Y_{\lm}.
\end{aligned}
\end{equation}
 We want to isolate the in--going piece of the decomposition and set this as the boundary condition for our field. This can be achieved by seperating the Bessel J function into constituent Hankel functions,
\begin{equation}
    J_{\ell+\frac{1}{2}}( k_{\infty} r^{\rm out}_*) = \frac{H^{(1)}_{\ell+\frac{1}{2}}( k_{\infty} r^{\rm out}_*) +  H^{(2)}_{\ell+\frac{1}{2}}( k_{\infty} r^{\rm out}_*)}{2}.
\end{equation}
Using this we can isolate $\mathrm{W}_{\lm}$ into in--going and out--going pieces respectively as
\begin{equation}\label{eq:hankelsep}
\begin{aligned}
	\mathrm{W}_{\lm} &= I^{(p)}_{\lm} + R^{(p)}_{\lm} &\\
\end{aligned}
\end{equation}
with
\begin{equation}\label{eq:Hankel}
\begin{aligned}
   I^{(p)}_{\lm} &= (-i)^{\ell+m} \pi \text{Sign}( k_{\infty}  r^{\rm out}_*)  P_{\lm}(\cos\beta)\\
		&\qquad\qquad\times\sqrt{\frac{(2\ell+1) (\ell-m)!}{2k_{\infty}  r^{\rm out}_*(\ell+m)!}} 
			H^{(2)}_{\ell+\frac{1}{2}}( k_{\infty}  r^{\rm out}_*)\, ,\\
    R^{(p)}_{\lm} &= (-i)^{\ell+m} \pi \text{Sign}( k_{\infty}  r^{\rm out}_*)  P_{\lm}(\cos\beta)\\
		&\qquad\qquad\times\sqrt{\frac{(2\ell+1) (\ell-m)!}{2k_{\infty}  r^{\rm out}_*(\ell+m)!}} 
			H^{(1)}_{\ell+\frac{1}{2}}( k_{\infty}  r^{\rm out}_*)\, . 
\end{aligned}
\end{equation}
Notably, from these expression it can be shown that  $\lim_{r\to\infty} I^{(p)}_{\lm}$ is equivalent to Eq.~(17) of \cite{Vicente:2022ivh}.

The Klein--Gordon equation on a Kerr background does not separate in spherical harmonics, but in spheroidal harmonics. We can transform between both bases by using spherical--spheroidal mixing coefficients~\cite{Hughes:2000ssa}, which for a fixed spheroidicity $\xi$ (as is the case here) we denote by $c_{\alpha \ell m}(\xi)$. Then, a spheroidal harmonic is decomposed into spherical harmonics as 
\begin{equation}\label{eq:mixingcoeffs}
	S_{\lm}(\theta,\varphi,\xi) = \sum_{\alpha \geq |m|}^{} c_{\alpha \ell m}(\xi)Y_{\alpha m}(\theta,\varphi).
\end{equation}
 From this, we obtain the coefficients of the in--going plane wave component in terms of spheroidal harmonics as
\begin{equation}\label{eq:fullbc}
	\mathrm{J}^{(p)}_{\lm}= \sum_{\ell' \geq |m|}C_{\ell'\lm}(i k_{\infty}a) I^{(p)}_{\ell'm} \, .
\end{equation}

\subsection{Implementation of boundary conditions}
We begin impleneting the boundary conditions by setting the purely in-going condition on the horizon.  
Firstly one should note that radial Klein--Gordon equation~\eqref{eq:radialeq} can be mapped to the confluent Heun equation~\cite{Hortacsu:2011rr,Borissov:2009bj,Hui:2019aqm}. Thus, solutions that are regular at the horizon, and hence in--going, can be written in terms of confluent Heun functions. We use this to set the inner boundary conditions at a finite radius, away from the horizon,
$r^{\rm in}_{\rm BC}\approx 5M$~\footnote{
Integrating numerically very close to the BH horizon introduces numerical instabilities. However, evaluating the special functions at very large distances is computationally very expensive. We found a fair trade-off between these two issues by setting the boundary conditions at $r^{\rm in}_{\rm BC}\approx 5M$.}.
Next, we integrate numerically the equation outwards up to large radii for each spheroidal mode. At large radii we may then assume the solution takes the form of Eq.~\eqref{eq:asymptoticform}. From the numerical solution, we can then extract the coefficients of the piece in--coming from past null infinity, given by, $I_{\lm}$.

We now want to fix that asymptotically the 
coefficient of the in--coming piece from past null infinity corresponds to the in--going component of a plane wave, given by $J_{\lm}^{(p)}$~\eqref{eq:fullbc}. In order to achieve this, it is sufficient to re--scale the solution by a factor $J_{\lm}^{(p)}\sqrt{(r^{\rm out}_{\rm BC})^2 + a^2}/I_{\lm}$, in virtue of the linearity of the problem. This then fixes the solution of the scalar field in the full spacetime.

Individually, equations  Eqs.~\eqref{eq:Hankel} are divergent subsequences of Eq.~\eqref{eq:WLM}. However, one can see that by fixing $r^{\rm out}_{\rm BC}$ (choosing a cloud size) and analysing the asymptotic structure of the numerical solution, the $\phi{\lm}$ modes with $\ell > k_\infty r^{\rm out}_*$ are exponentially suppressed, yielding a converging sum 
\footnote{This can be seen explicitly as in the $M\rightarrow0$ limit, setting $I^{(p)}_{\lm}$ as the in--going piece one recovers $W_{\lm}$ as the full solution. Through the Bessel-J function, $W_{\lm}$ has precisely this exponentially convergence property.}.
The field solution we obtain represents a plane wave scattered by the BH potential within some ball of radius $r^{\rm out}_*$ (outside this ball the field sharply decays). Having a setup that converges in $\ell$ such as this allows for a direct method of calculating finite drag forces without the need for any cutoff schemes like those employed in previous works \cite{Traykova:2023qyv,Vicente:2022ivh, Hui:2016ltb, Lancaster:2019mde, Clough_2021}.

The goal here, however, is to calculate spin--induced dynamical friction effects, which we expect to saturate and not depend on cloud size.  Hence in practice, we take $r^{\rm out}_{\rm BC}=10^7$, recovering the same boundary conditions as \cite{Vicente:2022ivh}, and then sum to an appropriate number of $\ell$--modes such that the forces have converged. An analysis of the convergence in $\ell$ of the forces can be found in Appendix~\ref{sec:converg}.

\subsection{Forces from the scalar field}
%
Having obtained the solution for the scalar field, we can now calculate momentum transfers between the field and the BH. Due to asymptotic flatness, we are provided with three translational Killing vectors ($\xi^i_{\mu} = \delta^i_{\mu}$), from which we can define a rate of change of  linear momentum,
\begin{equation}
   F^i =  \frac{d P^i}{d t} =  - \lim_{r\rightarrow\infty } \int_{S_r}  T^{r i} r^2 d \Omega\, ,
\end{equation}
where it is important to note that, in the large $r$ limit,
\begin{equation}
    \lim_{r\rightarrow \infty} r^2 T_{r i} \rightarrow  \hat{x}^{i} r T_{rr}\,, 
\end{equation} 
given $\hat{x}^i$ as the triad of cartesian unit vectors. The expressions $\hat{x}^i$ can be found analytically in terms of spherical harmonics, meaning the integral over the sphere becomes
\begin{align}\label{eq:forcesxyz}
\begin{aligned}
    F^x =& \sqrt{\frac{4 \pi }{6}}   \lim_{r\rightarrow\infty }  \int_{S_r}  \left(Y_{1 -1 }- Y_{1 1}\right)  T_{r r} r^2 d \Omega,\\
    F^y =&  - \sqrt{ \frac{4 \pi }{6} }  i  \lim_{r\rightarrow\infty } \int_{S_r}\left(Y_{1 -1 }+ Y_{1 1}\right) T_{r r} r^2 d \Omega,\\
    F^z =& \sqrt{\frac{4 \pi }{3}} \lim_{r\rightarrow\infty }  \int_{S^2}    Y_{1 0 }   T_{r r}  r^2 d \Omega\, .
\end{aligned}
\end{align}
At large radii we are now only required to calculate the (rr) component of the stress-energy tensor,
\begin{align}\label{eq:Trr}
    T_{rr} =& \sum_{\ell,m}\sum_{\ell',m'} t_{\lm}^{\ell' m'} S_{\lm}S^*_{\ell' m'}\, ,\\
    t_{\lm}^{\ell' m'}  =&\frac{1}{2}\left( \partial_r \phi_{\lm}   \partial_r \phi^*_{\ell' m'}+  k_{\infty}^2 \phi_{\lm} \phi^*_{\ell' m'} \right) .
\end{align}
Expanding the spheroidal terms in Eq.~\eqref{eq:Trr} over spherical harmonics, one can then appeal to the standard expressions for the Wigner 3-j symbols \cite{Wigner1993} to compute the integrals in Eq.~\eqref{eq:forcesxyz} analytically.
Finally, we bring these integrals (including the sums over mixing coefficients) together with the numerical interpolants for $\phi_{\lm}(r)$ to obtain radially oscillating functions of the forces. 
Analysing the asymptotic form of Eq.~\eqref{eq:asymptoticform} it can be seen that, to leading order, the forces behave as
\be
\lim_{r\rightarrow \infty} F  = A+ B e^{-2 i  k_{\infty} r_* }+ C e^{2 i k_{\infty} r_*} + \mathscr{O}\left(\frac{1}{r}\right)\,,\label{eq:forceform}
\ee
where $A, B$, and $C$ are coefficients to be determined. Including higher order terms in $\mathscr{O}\left(1/r\right)$ we perform a simple linear regression to extract the $A$ coefficients. These are then the values of $\{F^x, F^y, F^z \}$ for a given point in $\{\mu,a,v,\beta\}$ parameter space. 
Finally, we apply a rotation to the cartesian force vectors to obtain the drag, Magnus and lift forces,
\beq
F_D^{(W)} &=& F^z \cos \beta -F^y \sin \beta  ,\label{F_dragw}\\
F_M^{(W)} &=& F^x,\label{F_magnusw}\\     
F_L^{(W)} &=& F^y \cos \beta  + F^z \sin \beta\, . \label{F_liftw}
\eeq

\section{Results}\label{sec:results}
We have evaluated the aerodynamic forces \eqref{F_dragp}--\eqref{F_liftp} and \eqref{F_dragw}--\eqref{F_liftw} semi--analytically as described above in a grid of Chebyshev points. The grid then covers the parameter space given by the velocity $v$, the dimensionless spin $a/M$, and the incidence angle $\beta$. From these points, we build a Chebyshev interpolant so that we can explore parameter space efficiently. Convergence of our results with the size of the cloud and details of the implementation of the interpolant (including the ranges in parameter space used to build them) are given in Appendix~\ref{sec:converg}. The data used to build the interpolants can be found in~\cite{dyson_2024_10604046}.

\subsection{Comparison with previous works}\label{sec:comparison}
\begin{figure}[tbh!]
    \centering
    \includegraphics[width = 1\columnwidth]{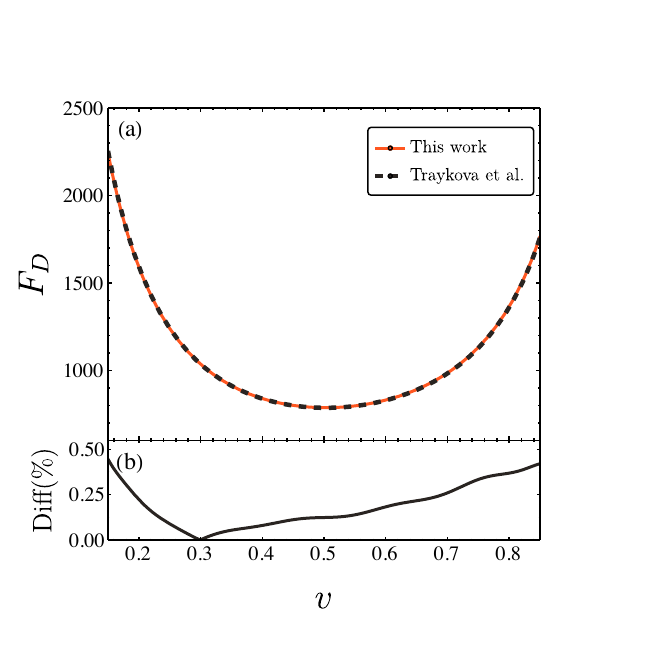}
    \caption{ \textbf{Top:} Drag force \eqref{F_dragp} for particle-like environments, at zero spin and almost head--on $\beta = 0.1$ compared with the analytical predictions obtained from Eq. (23) of~\cite{Traykova:2023qyv}.
    \textbf{Bottom:} Percent level of the relative difference between the values found following our method and the analyitcal expression of~\cite{Traykova:2023qyv}. It is clearly seen that the relative difference between our results and the analytical predictions is always sub--percent. 
    }
    \label{fig:Comparison_Traykova}
\end{figure}
\begin{figure}[tbh!]
    \includegraphics[width = 1\columnwidth]{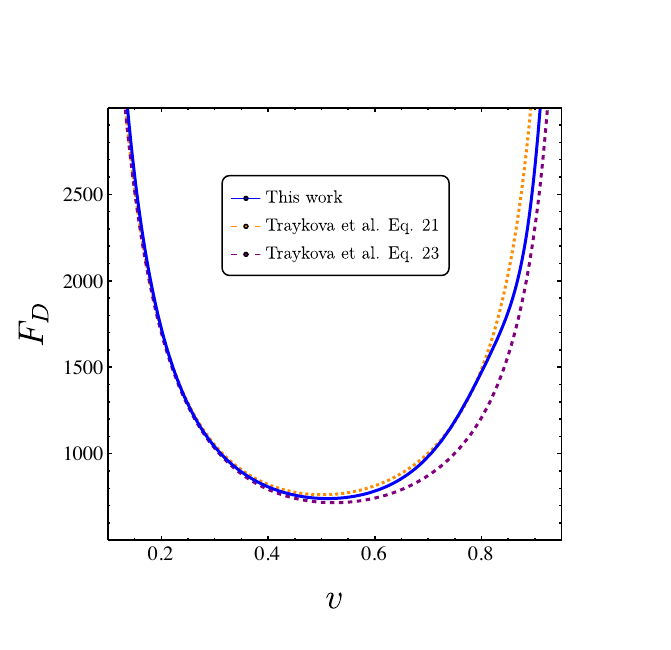}
    \centering
    \caption{Drag force \eqref{F_dragw} calculated for wave-like medium, summed to $\ell_{\rm max}=15$ with $\mu = 0.2, \beta =0.001$ and $a=0.0001M$. Note here that Eqs. (21) and (23) have been evaluated explicitly as given in~\cite{Traykova:2023qyv} with $R = 4000$. Here we find that the particle and wave approximations from \cite{Traykova:2023qyv} bound our values for the drag over the majority of the domain.} 
    \label{fig:Comparison_plot_fields}
\end{figure}
\begin{figure}[tbh!]
    \includegraphics[width = 1
\columnwidth]{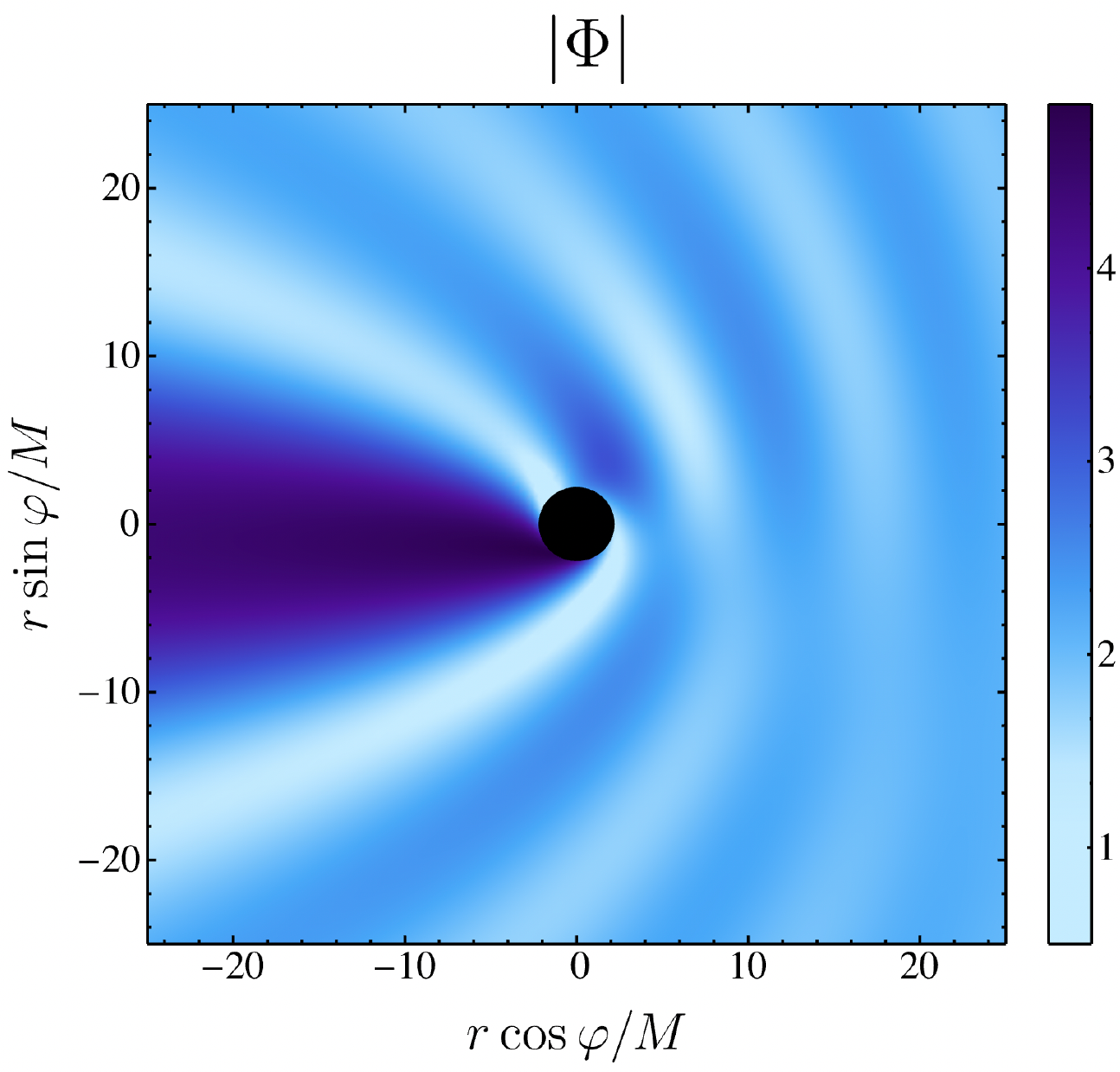}
    \centering
    \caption{Equatorial slice of the magnitude of the scalar field $\lvert \Phi\rvert $, where $\mu M = 0.2$, over a region close to the BH. Here $(r,\varphi)$ denote Boyer-Lindquist coordinates. The BH moves towards the right with $v = 0.8$, and its spin is along the axis pointing out of the plain, in an anti-clockwise direction, with magnitude $a=0.9M$. We observe a wake of higher field density forming behind the BH, and clear wave--like patterns. Moreover one can notice that due to the spin of the BH, reflection along the $\varphi=\{0,\pi
    \}$ axis is no longer a symmetry of the problem.
    }
    \label{fig:wavedensityplot}
\end{figure}
Our ultimate goal is to characterize the spin effects on dynamical friction. However it is important to first check that our semi--analytical set--ups are consistent with previous results found for BHs with no spin, but in the fully relativistic regime~\cite{Vicente:2022ivh, Traykova:2021dua, Traykova:2023qyv}. For a particle--like medium, analytical formulas for the dynamical friction force (for Schwarzschild BHs) 
that allow for a direct comparison with our work were given in Eq.(23) of~\cite{Traykova:2023qyv}. In Fig.~\ref{fig:Comparison_Traykova} we show that the relative difference between the drag force \eqref{F_dragp} that we obtain for non--rotating BHs and the analytical calculation is below $1\%$. The good accuracy of the comparison serves as a benchmark of our semi--analytical procedure discussed above. 

For a wave--like, scalar-field environment, when the wavelength of the field is comparable to the size of the BH $ \gamma \mu M \sim \mathscr{O}(1)$, the analytical work carried out for Schwarzschild BHs in Ref.~\cite{Vicente:2022ivh, Traykova:2023qyv} is only an approximation. In particular, Ref.~\cite{Traykova:2023qyv} found that the best match to numerical results was obtained by combining the scattering forces from the limit in which the scalar field is very light, $\gamma\mu M \ll 1$, but the accretion effects in the limit of particles (or very heavy scalars), $ \gamma \mu M \gg 1$. In Fig.~\ref{fig:Comparison_plot_fields} we observe that our results for the drag force \eqref{F_dragw} (the blue line) are compatible and bracketed by the different approximations discussed in Ref.~\cite{Traykova:2023qyv}. In particular, for this comparison we have set $\mu=0.2$ (which is the mass that we cover parameter space with),  giving $\gamma \mu \in (0.2,0.45)$  on the range $v\in(0.1,0.9)$. These values of  $\gamma \mu$ lay outside the validity of the approximations of \cite{Vicente:2022ivh,Traykova:2023qyv}. In particular one should note the non-uniform behavior for our results near $v=0.7$ in Fig.~\ref{fig:Comparison_plot_fields}. For this choice of $\mu$ it is approximately at this velocity that the wavelength of the field in the BH frame becomes comparable to the length scale of the innermost stable circular orbit.

Finally, it is informative to observe the scalar field density on an equatorial slice, when the BH is moving in a direction orthogonal to its spin, as shown in Fig.~\ref{fig:wavedensityplot}. For Schwarzschild BHs, we know that the field develops a cylindrically symmetric wake or an overdensity behind, that ultimately causes the drag force. Here we show for a rapidly spinning BH that a wake develops, and moreover that the field structure is asymmetric (c.f. Fig.1 in~\cite{Vicente:2022ivh}). While the wake behind the BH gives a visual interpretation for the origin of the drag force, this asymmetry is directly responsible for the Magnus force.

\subsection{Exploration of parameter space}\label{ssec:parameterspace}

\begin{figure*}[ptbh!]
    \centering
    \includegraphics[width = 1.0\textwidth]{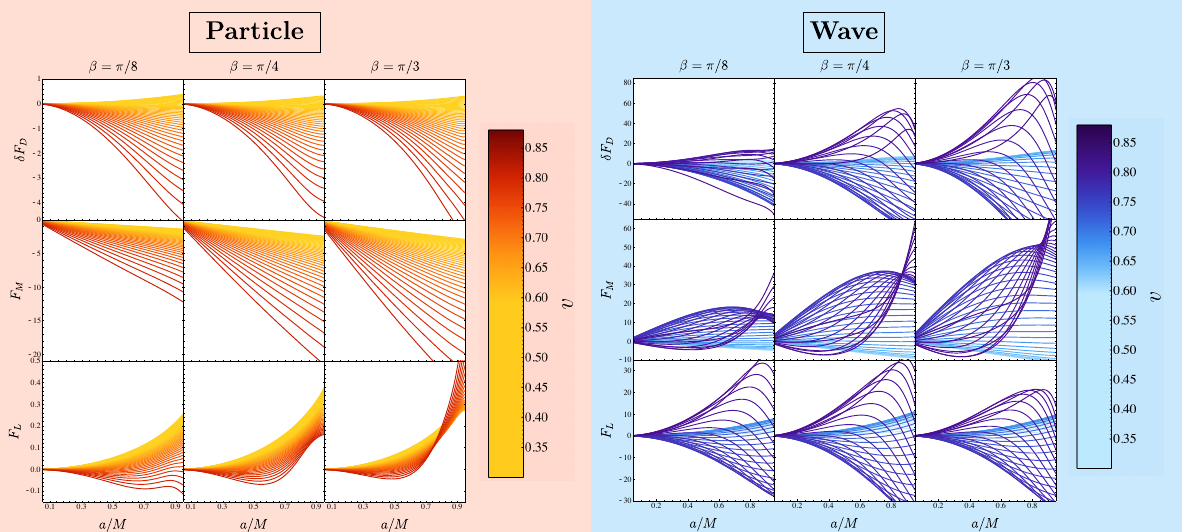}
    \centering
    \caption{ \textbf{Left:} The drag, Magnus and lift forces for different BH velocities and spin, in particle environments. Results are scaled by $\rho M^2$. We evaluate the forces at $\beta = \pi/8, \pi/4, $ and $ \pi/3$. In the low--velocity regime, the Magnus force is negative and there is a clear negative linear scaling with spin for all values of $\beta$ tested. On the other hand, the draf and lift forces at small velocities are all positive and increase with spin. \textbf{Right:}  Spin scaling for the variation in drag, Magnus, and lift forces for the wave setup. Here all quantities are evaluated equivalently to the particle case. Notice how in the low-velocity regime the spin scalings are the same as for the particle setup.
    } 
    \label{fig:spin_plot}
\end{figure*}

Having compared our methods with previous results we now focus our attention on understanding the large parameter space covered by the Chebyshev interpolants. Our results are summarized in Figs.~\ref{fig:spin_plot}--\ref{fig:velocity_plot}. Firstly, we highlight that the overall variations of the drag with the spin and incidence angle are relatively small, and in particular, they are comparable to the other aerodynamic forces. For this reason, we decide to study carefully the variation in drag $\delta F_D$, defined as 
\begin{equation}
\delta F_D(a,v,\beta) = F_D(a,v,\beta) - F_D(0,v,\beta) \, .
\end{equation}
As shown in Appendix~\ref{sec:converg}, this quantity does \emph{not} depend on the cloud size, making it a more suitable quantity to track than the drag force itself, which grows logarithmically with the size of the cloud.

\subsubsection{Dependence on spin}

Figure~\ref{fig:spin_plot} shows the dependence of the aerodynamic forces on the spin. The panel on the left (right) refers to a particle (scalar-field) environment. We focus on a scalar field with mass $\mu M=0.2$, such that as the velocity increases we can study the transition from wave optics to the regime where the scalar field begins to probe the strong--field substructure of the spacetime.

Due to the logarithmic divergence in the drag force, the variation on drag is a small effect. In particular, the magnitude of the drag force for parameters of Fig.~\ref{fig:spin_plot} is 3 orders of magnitude larger
than $\delta F_D$. Thus the plot shows the small variations of the drag with the BH spin, with an approximately quadratic dependence, at low spins.
We find that the Magnus force for particle-like environments is always negative. It is an {\it anti-Magnus} force, in reality, consistent with a post-Newtonian analysis in Ref.~\cite{Cashen:2016neh}. Our results disagree with the conclusions of Ref.~\cite{Costa:2018gva} regarding the sign of the force, but we note a word of caution of footnote~\ref{footnote_costa}, and also of Ref.~\cite{Costa:2018gva}, suggesting that force estimates from fluxes may yield incorrect values. 
Evidently from the figure, the Magnus force grows linearly with the spin of the BH for a particle--like medium. This is also the case for a wave--like medium, but only for low velocities. Note also that for scalar-field environments, the Magnus force can be positive even at low spins.
Since the Magnus force is ultimately a consequence of frame dragging, and the horizon frequency is linear in the BH spin, it is natural to expect that the Magnus force, at least in the low spin, low velocities regime, becomes linear in $a/M$.

Both the lift and the variation in the drag force, on the other hand, exhibit a quadratic scaling with the BH spin at low spins and low velocities. As can be seen from the construction of the aerodynamic forces in coordinates~\eqref{eq:forcesxyz}, the lift will ``compete'' with the drag force. Therefore we observe that they share some relevant features, such as the spin scaling. At large spins and velocity, the drag force can be enhanced or suppressed (relative to the non-spinning case).

Focusing on the high--velocity scenarios (represented by lines with a darker color), we observe that the same generic features as in the low velocity case persist for a particle--like medium, albeit with corrections that become more important for higher spins (see e.g. the change of sign in $F_L$). On the other hand, for a wave--like medium, once the velocity becomes $v \sim 0.7$ the forces begin to oscillate, introducing new features that are not present in the weak field. These are a consequence of the wave--like nature of the scalar field, which at $v = 0.7$ has a wavelength $\lambda_C \sim 10 M$, comparable to the length scale of the ISCO. We leave a detailed characterization of the strong field effects for future work. 

\subsubsection{Dependence on incidence angle }

\begin{figure*}[ptbh!]
    \centering
    \includegraphics[width = 1.0\textwidth]{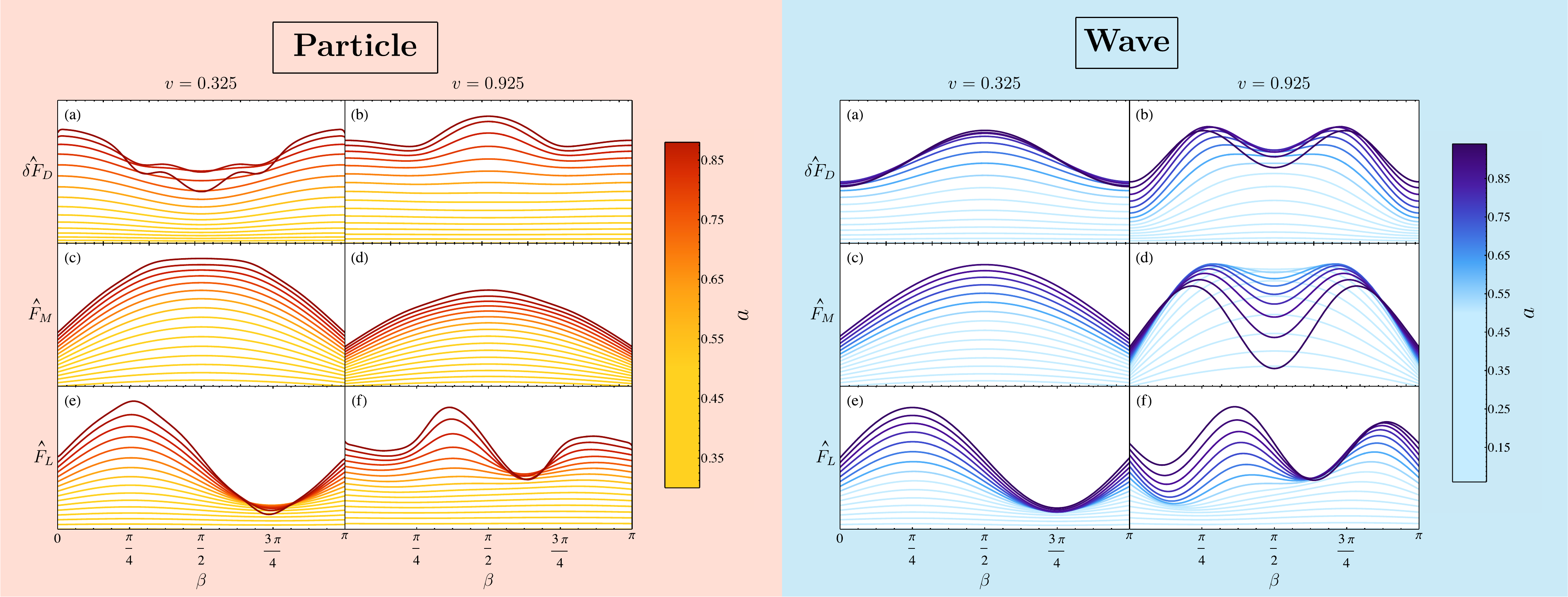}
    \centering
    \caption{
    \textbf{Left:} dependence of forces for particle-like environments with incidence angle $\beta$ for two different BH velocities, $v=0.325$ and $0.925$. In the regions of parameter space tested, the Magnus force is always maximised when $\beta=\pi/2$ (i.e. when the spin and velocity vectors are orthogonal). Note also the development of additional maxima and minima that arise in the lift force.  \textbf{Right:} same, for scalar-field environments, here with $\mu M=0.2$. Features are similar to those of particle environments, but now the Magnus force also develops additional maxima and minima in the high velocity/spin regime.} 
    \label{fig:joy_plots}
\end{figure*}

Fig.~\ref{fig:spin_plot} also shows that the aerodynamic forces have an interesting structure depending on the incidence angle $\beta$. 
To study this further, we show the overall angular structure of the forces in Fig.~\ref{fig:joy_plots}. In order to better understand the structure of the $\beta$ dependence we perform a re--scaling of the forces of the form,
\begin{equation}
\hat{F_i}(\beta) = \frac{F(a_i,v_0,\beta)}{\max_{\beta} F(a_{max},v_0,\beta)} + \frac{i}{c_0},
\end{equation}
where $v_0 \in \{0.325,0.925\}$, $i$ indexes the individual lines of different spin values in the subplots, and $c_0$ is some constant chosen to provide a clear separation between beta scalings for different spin values in each particular subplot. 
This arbitrary shift and scaling of the forces in a given subplot means the magnitude of individual lines on a single plot may be compared, however, we do not focus on the absolute values between different subplots, but instead on the angular pattern.
It is evident that at low velocities the Magnus force has a clear $F_M\sim \sin\beta$ behavior, whereas $F_L \sim \sin(2\beta)$. The variation in the drag has a more complex structure. Worth remarking however is that the overall value in $\delta F_D$ is much smaller at low velocities than it is at high velocities. 

We find that the angular structure changes significantly at high spins and at high velocities: the lift force, in particular, develops an additional set of maxima and minima between which new roots arise. Interestingly, we have found these roots asymptote to $\pi/4$ and $3\pi/4$ in the limit $a,v\to 1$ (which is the point where the lift forces are maximised in the low-velocity limit). For the case where the wavelength of the medium becomes comparable with the size of the BH, the angular structure of the Magnus force and $\delta F_D$ also change. Thus, the angle at which each of the aerodynamic forces is maximized depends sensitively on the BH velocity and spin.

\subsubsection{Dependence on velocity}
%
\begin{figure*}[tbh!]
    \centering
    \includegraphics[width = 1.0\textwidth]{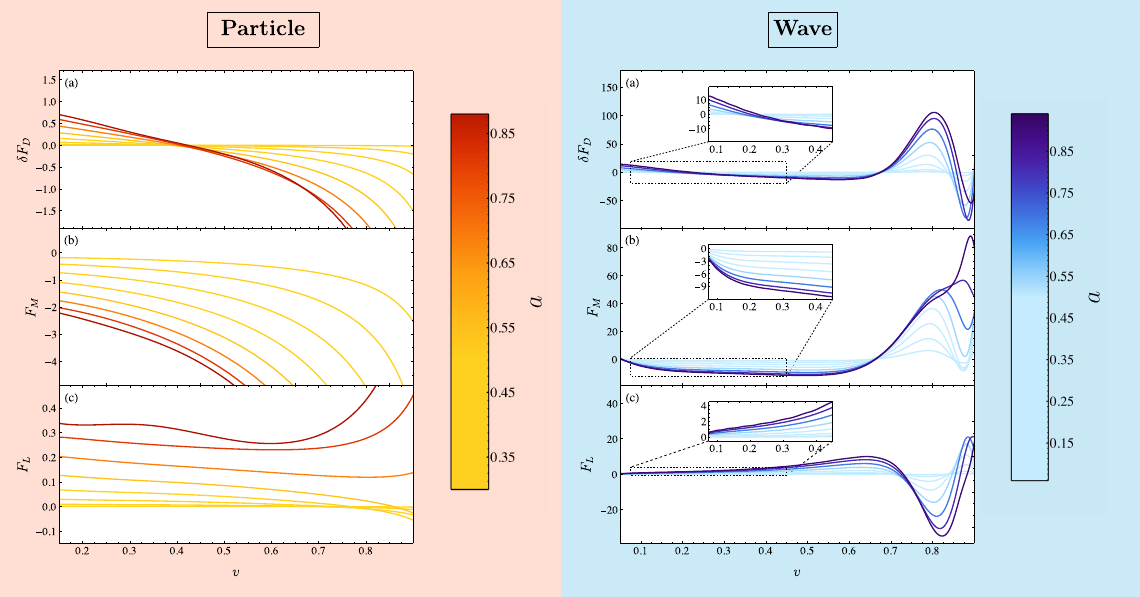}
    \centering
    \caption{\textbf{Left:} Velocity dependence of the spin aerodynamics forces in the particle setup. Here we show the velocity scaling for a family of differing spin values (increasing in spin with darkness) evaluated at $\beta   =\pi/3$. We find a transition from positive to negative variation in the drag at relatively low velocities with negative Magnus and positive lift forces. \textbf{Right:} Velocity scaling of the spin aerodynamics forces for the wave setup. Here all quantities are evaluated equivalently to the particle case, the insets run on the range $v\in(0.05,0.4)$ and show explicitly in the low-velocity limit that the particle and wave cases have the same qualitative behavior. Returning to the full plot on the range $v\in(0.05,0.9)$ one again sees a transition in the sign and an order of magnitude increase in the forces as the wavelength moves to scales relevant for probing the strong field structure. 
    } 
    \label{fig:velocity_plot}
\end{figure*}
Finally, we characterize the dependence of the aerodynamic forces on the velocity in Fig.~\ref{fig:velocity_plot}, where the forces are shown at an angle $\pi/3$.
A cursory inspection of this figure confirms the qualitative agreement between the particle and wave setups. We observe the same characteristic sign change in the drag variation early in the velocity scaling. One also finds at low velocities a lift force that is positive for both setups. Additionally, at low velocities, we obtain an anti-Magnus force driving the BH in the opposite direction one would expect from the intuition provided by classical aerodynamics. 

As velocity increases we also observe an order of magnitude increase in the forces for both the particle and wave mediums (except for the particle lift force). This can be thought of as a consequence of the BH interacting with an increasingly dense environment due to length contraction. In the wave set--up we find that as the velocity passes through $v\simeq 0.7$ there is a change in the sign of the forces. In simulating these scaling to higher velocities than is shown here ($v=0.95$), it can be seen that this change in sign is characteristic of novel oscillatory effects that arise in the high--velocity limit, as the system transitions from the wave optics to geometric optics regimes. An in--depth analysis of this transition and the geometric optics limit of the scalar field is left as a task for future work.

\subsection{Polynomial fits}\label{ssec:polynomialfits}
For convenience, we provide polynomial fits to the aerodynamic forces in the weak field regime, which describe well our numerical results. The procedure developed here follows closely the construction of hyperfits for a ringdown model~\cite{Cheung:2023vki}. We provide fits for the range $a/M \in [0.1, 0.6]$ and $v \in [0.15, 0.5]$ (for a particle--like medium), and $a/M \in [0.04, 0.6]$, $v\in[0.05, 0.5]$ (for a wave--like medium with $\mu M = 0.2$). In order to do so, we first extract the expected scaling of the forces with spin and angles, e.g., we define $\Tilde{F}_M = F_M / (a\sin\beta)$, and similarly $\Tilde{F}_L = F_L / (a^2 \sin(2\beta)$ and $\Tilde{\delta F}_D = \delta F_D / a^2$. We perform a linear regression fit on the rescaled forces for a polynomial of degree $N$ in the variables $\{a, \sqrt{v}, \cos\beta\}$. We fix the degree of the polynomial once the goodness of fit measure achieves a certain minimum value. We use as a goodness of fit the adjusted R squared measure, $\bar{\mathbf{R}}^2$, defined as  
\begin{equation}
    \bar{\mathbf{R}}^2 = 1 - \frac{n-1}{n-p}\mathbf{R}^2 \, ,
\end{equation}
where $n$ is the total number of points used for the fit ($1024$ for a particle medium, and $4000$ for a wave medium), $p$ is the number of parameters in the model, and $\mathbf{R}^2$ is the usual linear regression coefficient. We compute this for polynomials of different degree, and stop at the degree $N$ for which $\bar{\mathbf{R}}^2 \geq 0.999$~\footnote{Extracting the structure in the variations of the drag force proves to be more challenging, so we relax the accuracy requirement to $\bar{\mathbf{R}}^2 \geq 0.99$.}.

Finally, since a polynomial of, e.g. degree $N=3$ in $3$ variables contains $p=20$ terms, we would like to reduce the number of terms included in the polynomial. To do this, we rank the contribution of each of the terms by computing the mismatch of the model once we do not include that given term. Then, we add one by one each term, in the order in which they were ranked, until the previous tolerance limit is achieved.

The fits obtained are given by\footnote{Here we reiterate that a positive $\delta F_D$ refers to a variation that enhances the drag force causing the BH to slow down faster. Algebraic signs follow conventions in Fig.~\ref{fig:forces}. A positive $F_M$ is one in which the force is acting as in the standard classical fluid mechanics. A positive $F_L$ force is one which is pushing the BH upwards along the positive z-axis in the coordinate we define in sec.~\ref{sec:particle_medium}.}:
\begin{widetext}
\begin{equation}\label{eq:Analytical_Scalings}
    \begin{aligned}
        \delta F_D^{(P)}  =& a^2 \Bigl(1.11 - 3.26 v +0.74 a-1.07 \sqrt{v}a-0.59 \cos\beta +1.38 \sqrt{v} \cos\beta\Bigr) \, , \\
        F_M^{(P)}  =& a \sin\beta\Bigl(2.55-31.81 \sqrt{v}+69.81 v-59.15 v^{3/2}-0.12 a\sqrt{v} \cos\beta  \Bigr)\, , \\
        F_L^{(P)} =& a^2 \sin 2\beta\Bigl(0.82-2.60 \sqrt{v}+5.06 v-3.79 v^{3/2}\Bigr.\\
        \Bigl.&+0.28 a+1.28 v a-0.26 \sqrt{v} a\cos\beta +0.11a \cos\beta-1.22 \sqrt{v} a-0.02 \cos\beta+0.04\sqrt{v} \cos\beta+0.23 \sqrt{v} a^2\Bigr) \, , \\ 
        \delta F_D^{(W)} =& a^2 \Bigl(-16.45-30.84 \sqrt{v}+316.40 v-296.10 v^{3/2}+5.44 \cos^2\beta-23.92 \sqrt{v} \cos^2\beta+12.25 \sqrt{v} a^2-9.16 a^2\Bigr) \, , \\
        F_M^{(W)} =& a \sin\beta\Bigl(36.45-241.9 \sqrt{v}+415.8 v-248.6 v^{3/2}\Bigr) \, , \\
        F_L^{(W)} =& a^2 \sin 2\beta\Bigl(-1.023463+58.2938 v-165.9078 v^{3/2}+150.6706 v^2\Bigr) \, .
    \end{aligned}
\end{equation}
\end{widetext}
For the Magnus force in particle environments, we can compare to the post--Newtonian based calculations carried out in Ref.~~\cite{Cashen:2016neh}. The first correction, in the zero--velocity limit, in that case, would be $\sim -2.5$. Evaluating our fit at the lowest velocity available $v = 0.15$, and low spin, returns $F_M \sim -2.73$, in close agreement with the results of~\cite{Cashen:2016neh} (we remind the reader that low velocities $v < 0.15$ were not available to construct the fit, and hence evaluating the fit provided above at exactly zero velocity would result in large extrapolation errors).

\section{Discussion}\label{sec:discussion}

In previous sections we studied and characterized the way in which the dynamics of BHs moving in a nontrivial environment is modified due to BH spin. We can identify two candidate scenarios for which the spin--dependence of the aerodynamic forces leave an observational imprint: (i) isolated, supermassive BHs moving through the galactic medium would have their trajectory curved due to the Magnus and lift forces. Therefore observing a curvature on the trail left by such a BH would be a smoking gun of the interplay between its spin and the medium through which the BH moves. (ii) Extreme mass ratio inspirals (EMRIs) consist of binary systems formed by a solar mass BH and a supermassive BH. These are a promising source of gravitational waves in the LISA band~\cite{LISA:2022yao}, where they would emit long-lasting signals that can be modelled very accurately. The presence of a medium surrounding the central supermassive BH would change the orbital dynamics of the companion, resulting in a modified gravitational wave emission. We consider these separately.

\subsection{Isolated BHs}

Supermassive BHs growing in active galactic nuclei (e.g. through consecutive mergers~\cite{Fabj:2020qqc}) can be ejected from said nuclei at relativistic velocities~\cite{Bekenstein:1973zz,saslaw1974gravitational, Volonteri:2002vz,Gonzalez:2006md,Merritt:2004xa, Lousto:2011kp}. As it is kicked out, it carries with it a significant amount of gas, that forms a wake behind it. This leaves trails that can be observed in the electromagnetic spectrum~\cite{Civano:2010es,vanDokkum:2023wed, Ogiya:2023ljz}~\footnote{Notice that some recent works argue that the observed trail could be due to a bulgeless, edge-on galaxy~\cite{Almeida:2023flb, Almeida:2023thu, vanDokkum:2023egp}.}.
 
Consider a simplified set-up where the BH spin and its velocity are orthogonal, i.e., $\beta = \pi/2$. After a time $T$ has passed, a spinning BH moving through a medium would be displaced by a distance $d = F_M T^2 / (2M)$ with respect to a spin--less BH (or one moving in vacuum), due to the Magnus force. If trail formation is observed behind it, at an angular distance $D_A$, the angle that would be needed to resolve in order to observe this deviation is $\theta = d / D_A$, including some significant numbers yields
\begin{equation}
    \lvert \theta \rvert \sim 8\times 10^{-20} \mathrm{arcsec} \Quantity{n}{}{\mathrm{cm}^{-3}} \Quantity{T}{}{\mathrm{kyr}}^2\Quantity{M}{10^8}{M_\odot}\Quantity{\mathrm{Gyr}}{D_A}{} \, ,
\end{equation}
where we have assumed that the BH moves with $v=0.15$, and $a=0.5M$. Therefore, hydrogen number densities of $n \sim 10^{17} \mathrm{cm}^{-3}$ over a scale of $T \sim kyr$ (equivalently, spanning a range of $\sim 45 \mathrm{pc}$) are necessary in order to observe a significant deviation, when observed at approximately redshift $z=1$. Dark matter spikes \cite{Gondolo_1999,Bertone_2005}, for example, are capable of clustering to much higher densities (close to the density of water, $n \sim 10^{23} \mathrm{cm}^{-3}$) in the vicinity of supermassive BHs, but the density of the halo decreases sharply at larger distances. 

\subsection{EMRI aerodynamics }

EMRIs are a very interesting prospect to learn about the geometry and the astrophysical environment very close to supermassive BHs~\cite{Babak:2017tow}. The smaller object, typically with a mass a million times smaller than its companion, can orbit for years at distances smaller than $10$ Schwarzschild radius of the central BH. The prospect of observing gravitational waves emitted from these systems with the space interferometer LISA motivates studying the impact of possible environments (such as gas, accretion disks or dark matter) on their emission patterns~\cite{Barack:2018yvs, LISAConsortiumWaveformWorkingGroup:2023arg}. 

 In addition to gravitational wave emission, if there is matter in the vicinity of the central BH, it will create further dissipative channels that will affect the companion's orbital motion, whereby changing the trajectory, would also modify the gravitational wave emission. Several studies in recent years have discussed that this is a potential opportunity to learn about said environments from the GW observations~\cite{Caputo:2020irr, Speeney:2022ryg, Speri:2022upm, Cole:2022yzw, Cardoso:2022fbq, Cardoso:2021wlq, Cardoso:2022whc, Kuntz:2022juv}. Here we consider whether the coupling between the spin of the secondary BH and the environment (through variations in the drag force, or the Magnus and lift forces, for instance), could have an impact. 

The spin--dependent aerodynamic forces that we have discussed scale roughly as 
\begin{equation}
    F_X \equiv \mathcal{C} q^2 \frac{a}{M} f(v,a,\beta) \, , \quad \mathcal{C} = \rho M^2_{\rm SMBH} \, , 
\end{equation}
where $q = M / M_{\rm SMBH}$ is the mass ratio between the mass of the secondary, $M$ and of the primary, $M_{\rm SMBH}$, $f(v,a,\beta)$ is a dimesionless quantity, of roughly order $1$ (see the leading term in Eq.~\eqref{eq:Analytical_Scalings}, and $\mathcal{C}$ is the (also dimensionless) number that controls the overall strength of the interaction. 

The spin--dependent aerodynamic forces on the smaller body therefore appear at the same formal order as the effective Matthison--Papapetrou--Dixon (MPD) force~\cite{Mathisson:1937zz,Papapetrou:1951pa,Dixon:1974xoz} due to the secondary spin coupling to the background geometry. However, unlike the MPD force the spin--dependent aerodynamic forces can act dissipatively on the binary, secularly changing the constants of motion and appearing at leading order in the post-adiabatic expansion of the EMRI dynamics in terms of the mass ratio. Moreover, the dimensionless quantity that controls their effect is not necessarily very small ($\mathcal{C}\sim 10^{-6}$ for environments with the density of the water, and $M_{\rm SMBH} = 10^6$). This raises the prospect of the aerodynamic forces being the dominant effect due to the spin of the smaller object, potentially improving its detectability.
A full assessment of the impact of the coupling between the spin and the environment in EMRIs will, however, require an in depth treatment of more realistic scenarios (including e.g. understanding the aerodynamic forces in generic orbits, and not in rectilinear motion). Furthermore, in any scenario where the spin-induced aerodynamic forces are significant, the regular component of the dynamical friction will be even bigger. A firm handle on the modelling uncertainties in the environment will therefore be paramount for any robust inference.

\section{Conclusions}\label{sec:conclusions}
In this work, we have studied the impact of BH spin on dynamical friction in the fully relativistic regime. When the BH spin does not point along its direction of motion, cylindrical symmetry of the problem is broken, and two additional forces arise. These we dub Magnus and lift forces. Therefore the motion of a spinning BH through a medium will generically be curved out of its original plane of motion, as it decelerates. Besides, the spin-induced aerodynamic forces, one would also expect the interaction of the spin with the environment to induce an effective torque on the spin itself~\cite{Cashen:2016neh}. We have not studied this effect in this work. Nonetheless, the introduced methods are amenable to such an enterprise.

We have considered two different kinds of environments, representing very distinct physical regimes: on the one hand, a medium composed of collisionless massive particles, and on the other hand, an ultralight scalar field, in the regime where wave and strong field effects are most important. In the first case, we provide a semi--analytical calculation of these forces using the properties of scattering timelike geodesics on a Kerr background. In the second case, we have implemented analytically the boundary conditions corresponding to a plane wave scattering off an angle of a Kerr BH, and solved the relevant equations numerically. Our results agree with previous analytical (and also numerical) calculations in the fully relativistic regime in the case where the BH spin is aligned with its velocity~\cite{Vicente:2022ivh, Traykova:2023qyv}, as shown in Sec.~\ref{sec:comparison}. 

We have evaluated the forces on a grid of points in parameter space, which we share publicly in~\cite{dyson_2024_10604046} and use to build rapid Chebyshev interpolants. From these we have learned the general features of the aerodynamic forces, including its spin scaling, angular structure, and dependence on the velocity of the BH. We also provide polynomial fits that are accurate in the regimes where the spin and the velocity of the BH are small or intermediate. Exploring in full detail the behaviour of the aerodynamic forces in the strong field regime, e.g., for very rapidly rotating BHs, is left for future study.

Finally, we have also discussed two classes of astrophysical systems in which the spin dependence of the dynamical friction could be of relevance. We observe that if runaway supermassive BHs move through dense enough environments, they could form curved trails. More interestingly, we argue that for an EMRI system the aerodynamic forces could be the dominant consequence of the spin of the smaller object. Further exploration of the spin--dependent aerodynamic forces in generic orbits, as well as its impact on gravitational wave emission, will be necessary to understand the full implications for EMRI observations. 

In the process of carrying out our work on the spin aerodynamic forces, we became aware of a concurrent endeavor to study the gravitational Magnus effect from the perspective of Numerical Relativity  \cite{Wang:2024cej}. This work makes use of the \texttt{GRDzadzha}
code--base \cite{ Aurrekoetxea:2023fhl,Andrade:2021rbd,Clough:2015sqa} to calculate the Magnus force up to high spin and intermediary relativist velocities (with the spin vector fixed to be perpendicular to the velocity). We see promising agreement between our results but leave a full comparative analysis as a task for future work.

\acknowledgments
We are indebted to Roberto Carlos for providing the initial inspiration to study this effect. 
The authors thank Katy Clough and Dina Traykova for insightful conversations on the Black Hole Magnus effect.
We also acknowledge insightful discussions with Daniel J. D'Orazio, Lorenz Zwick, and David O'Neill and thank Hector O. Silva for bringing to our attention useful references. 
V.C.\ is a Villum Investigator and a DNRF Chair. We acknowledge financial support by the VILLUM Foundation (grant no. VIL37766) and the DNRF Chair program (grant no. DNRF162) by the Danish National Research Foundation. V.C. acknowledges financial support provided under the European Union’s H2020 ERC Advanced Grant “Black holes: gravitational engines of discovery” grant agreement
no. Gravitas–101052587. Views and opinions expressed are however those of the author only and do not necessarily reflect those of the European Union or the European Research Council. Neither the European Union nor the granting authority can be held responsible for them.
This project has received funding from the European Union's Horizon 2020 research and innovation program under the Marie Sklodowska-Curie grant agreement No 101007855 and No 101131233. We acknowledge financial support provided by FCT/Portugal through grants 
2022.01324.PTDC, PTDC/FIS-AST/7002/2020, UIDB/00099/2020 and UIDB/04459/2020.
This work makes use of the Black Hole Perturbation Toolkit. 
The Tycho supercomputer hosted at the SCIENCE HPC center at the University of Copenhagen was used for supporting this work.
\bibliography{ref}


\appendix

\section{Convergence}\label{sec:converg}

In this Appendix we provide additional numerical evidence for the convergence of our results. 

\begin{figure*}[ptbh!]
    \centering
    \includegraphics[width = .95\textwidth]{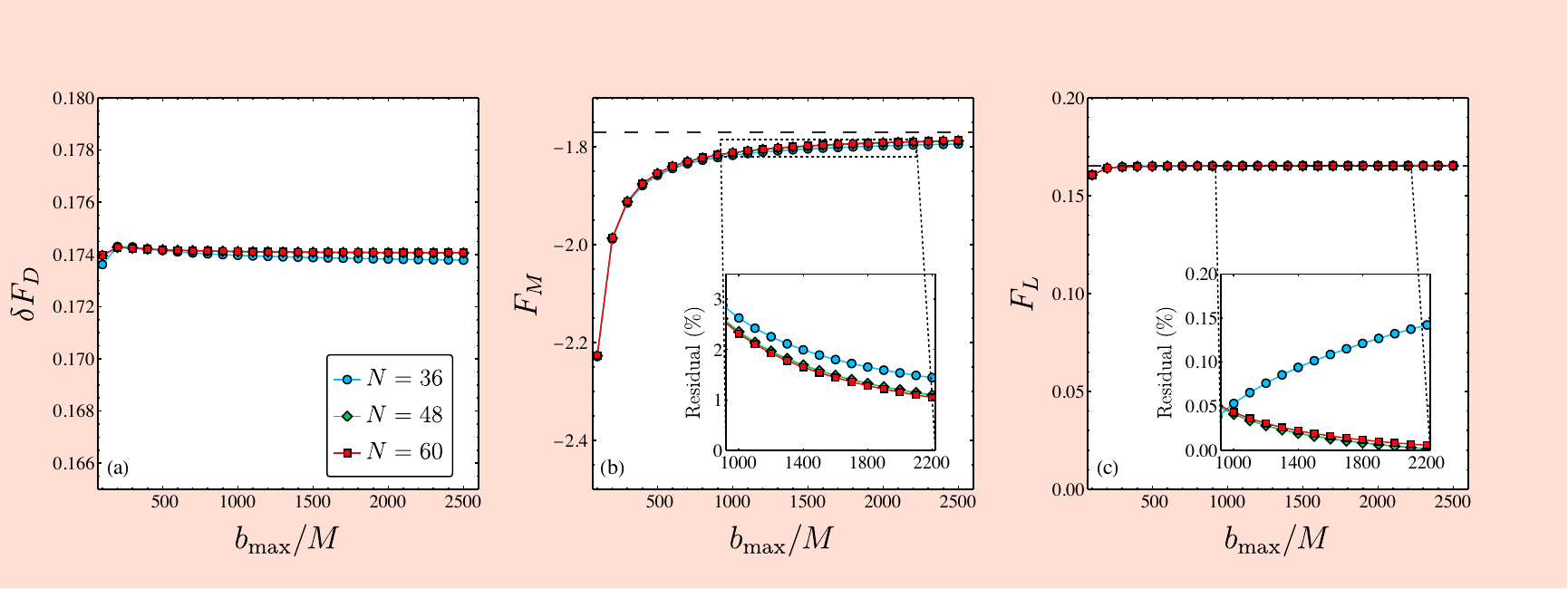}
    \caption{\textbf{(a):} Difference between the drag force evaluated at spin $a/M = 0.7$ and at very low spin, $a/M = 0.01$, keeping fixed the velocity $v = 0.3$ and the impact angle $\beta = \pi/4$, for different numerical resolutions (different colors, as shown in the label), as a function of the size of the cloud. We observe that this rapidly converges to a constant, showing that the spin-variation of the drag force is independent of the cloud size. \textbf{(b): } Scaling of the Magnus force with the cloud size, for a BH spin $a/M = 0.7$, and the rest of the parameters equal to the previous panel. The inset panel shows the residual between the asymptotic value (obtaining by fitting the highest resolution available to a series of the type $A + c_1 / b_{\rm max} + c_2 / b^2_{\rm max}$) and the data points, in percent level. This shows that the higher resolutions converge towards zero, achieving an accuracy of around percent level for the Magnus. \textbf{(c): } Same, but for the lift force.} 
    \label{fig:Convergence_test}
\end{figure*}

For the case of a particle--like medium, there are two factors to take into account: On the one hand, we numerically integrate over a disc in the impact plane. That numerical integration is performed on interpolants which use $N^2$ points. For each of those points the forces are computed analytically as described in the main text. Therefore, the only parameter controlling the numerical convergence is given by $N$. Fig.~\ref{fig:Convergence_test} shows the aerodynamic forces for three different resolutions. The insets for the Magnus and Lift forces show that increasing the number of points achieves convergence. 

Ultimately we want to extract the values of the forces for a cloud of a fixed size $b_{\rm max}$ where the quantities, $\delta F_D, F_M, F_L$ should \emph{not} depend on the cloud size (since they appear due to the strong field region near the BH). Fig.~\ref{fig:Convergence_test} shows that increasing $b_{\rm max} \geq 1000$ does not change the aerodynamic forces significantly, allowing us to safely perform the extraction at $b_{\rm max} = 1500$. The interpolant is constructed using the highest resolution available, given by $N = 64$. It uses $N_{\rm Cheb} = 16^3$ points distributed in the ranges $a/M \in [0.05, 0.9]$, $\beta \in [0.05, 1.52]$ (reflection symmetry is later enforced, but has been checked separately), and $v\in [0.15, 0.925]$.

\begin{figure}[ptbh!]
    \includegraphics[width = 1\columnwidth]{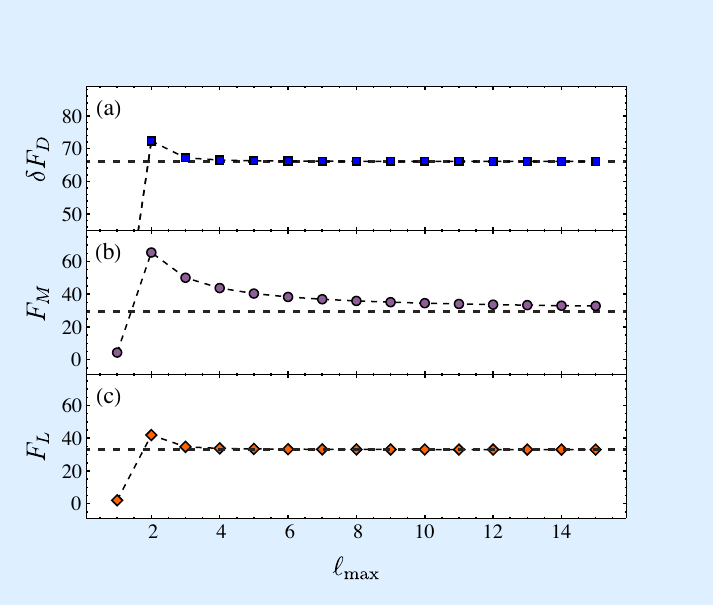}
    \centering
    \caption{
    Spin aerodynamic forces evaluated at spin $a/M = 0.95$, $v = 0.8$ and the impact angle $\beta = \pi/3$, as function of the $\ell_{max}$. We observe that this variation in drag and lift forces rapidly converges to a constant, whereas the Magnus has a slightly slower convergence rate just in the case of the particle medium.
    } 
    \label{fig:Converg_plot_fields}
\end{figure}

For the wave--like environment, there are also two convergence factors to consider. One being the accuracy requirements of the shooting method in numerically solving for the radial component of the scalar field. The other is the number of $\ell$-modes summed to obtain the result. The accuracy of the numerical solver was found to be extremely stable once an explicit Runge-Kutta method was implemented and so was set with a precision goal of 16 and allowed 32 digits of arbitrary precision arithmetic. The more subtle factor in convergence however is the $\ell$-mode sum. Following the same ethos as the cloud size in the convergence check for the particle case. We show in Fig.~\ref{fig:Converg_plot_fields} that for the case $\mu=0.2$ we have rapid convergence in our measured forces as a function of $\ell_{max}$ (which has an interpretation of cloud size here). In practice, we choose $\ell_{max} =15$ as an appropriately converged number of $\ell$-modes for the calculation of the nodes in the Chebyshev interpolant. We have also additionally taken the time to check that at the most extreme edges of our Chebyshev grid (around $a=0.9997, v = 0.95$) $\ell_{max} =15$ is also sufficient for convergence. The interpolant is constructed using $20$ nodes in the spin and angular directions, in the ranges $a/M \in [10^{-5}, 0.9997]$ and $\beta \in [10^{-5}, \pi-10^{-5}]$, and $40$ points in the velocity range given by $v \in [0.05, 0.95]$.

\end{document}